\documentclass[pra,aps,twocolumn,superscriptaddress]{revtex4}
\usepackage{graphicx}
\usepackage{times}
\usepackage{amsmath}
\usepackage{amsfonts}
\usepackage{amssymb}
\usepackage{latexsym}
\usepackage{color}
\usepackage{enumerate}
 \usepackage{algorithm}
 \usepackage{algorithmic}
 \usepackage{cases}
 \usepackage{bm}
\usepackage{comment}

\def\ket#1{\left|#1\right\rangle}
\def\bra#1{\left\langle#1\right|}
\def\avg#1{\left\langle#1\right\rangle}

\def\U#1{{\rm #1}}
\def\u#1{_{\rm #1}}

\begin{document}
\widetext
\title{Nuclear magnetic resonance spectroscopy with a superconducting flux qubit}
\author{Koichiro~ Miyanishi}
\email{miyanishi@qc.ee.es.osaka-u.ac.jp}
\affiliation{NTT Basic Research Laboratories, NTT Corporation, 3-1 Morinosato-Wakamiya, Atsugi, Kanagawa 243-0198, Japan}
\affiliation{Graduate School of Engineering Science, Osaka University,1-3 Machikaneyama, Toyonaka, Osaka 560-8531, Japan}
\author{Yuichiro~ Matsuzaki}
\email{matsuzaki.yuichiro@aist.go.jp}
\altaffiliation[Current Address: ]{ Nanoelectronics Research Institute, National Institute of Advanced Industrial Science and Technology (AIST), 1-1-1 Umezono, Tsukuba, Ibaraki 305-8568 Japan}
\affiliation{NTT Basic Research Laboratories, NTT Corporation, 3-1 Morinosato-Wakamiya, Atsugi, Kanagawa 243-0198, Japan}
\affiliation{NTT Theoretical Quantum Physics Center, NTT Corporation, 3-1 Morinosato-Wakamiya, Atsugi, Kanagawa 243-0198, Japan}
\author{Hiraku~ Toida}
\affiliation{NTT Basic Research Laboratories, NTT Corporation, 3-1 Morinosato-Wakamiya, Atsugi, Kanagawa 243-0198, Japan}
\author{Kosuke~ Kakuyanagi}
\affiliation{NTT Basic Research Laboratories, NTT Corporation, 3-1 Morinosato-Wakamiya, Atsugi, Kanagawa 243-0198, Japan}
\author{Makoto~ Negoro}
\affiliation{Quantum Information and Quantum Biology Division, Institute for Open and Transdisciplinary Research Initiatives, Osaka University}
\affiliation{JST, PRESTO, Kawaguchi, Japan}
\author{Masahiro~ Kitagawa}
\affiliation{Graduate School of Engineering Science, Osaka University,1-3 Machikaneyama, Toyonaka, Osaka 560-8531, Japan}
\affiliation{Quantum Information and Quantum Biology Division, Institute for Open and Transdisciplinary Research Initiatives, Osaka University}
\author{Shiro~ Saito}
\affiliation{NTT Basic Research Laboratories, NTT Corporation, 3-1 Morinosato-Wakamiya, Atsugi, Kanagawa 243-0198, Japan}

\begin{abstract}
We theoretically analyze the performance of the nuclear magnetic resonance~(NMR) spectroscopy with a superconducting flux qubit~(FQ).
Such NMR with the FQ is attractive because of the possibility to detect the relatively small number of nuclear spins in a local region ($\sim\mu$m) with low temperatures ($\sim$ mK) and low magnetic fields ($\sim$ mT), in which other types of quantum sensing schemes cannot easily access.
A sample containing nuclear spins is directly attached on the FQ, and the FQ is used as a magnetometer to detect magnetic fields from the nuclear spins.
Especially, we consider two types of approaches to NMR with the FQ.
One of them is to use spatially inhomogeneous excitations of the nuclear spins, which are induced by a spatially asymmetric driving with radio frequency~(RF) pulses.
Such an inhomogeneity causes a change in the DC magnetic flux penetrating a loop of the FQ, which can be detected by a standard Ramsey measurement on the FQ.
The other approach is to use a  dynamical decoupling on the FQ to measure AC magnetic fields induced by Larmor precession of the nuclear spins.
In this case, neither a spin excitation nor a spin polarization is required since the signal comes from fluctuating magnetic fields of the nuclear spins.
We calculate the minimum detectable density~(number) of the nuclear spins for the FQ with experimentally feasible parameters.
We show that the minimum detectable density~(number) of the nuclear spins with these approaches
is around $10^{21}$~/cm$^3$ ($10^8$) with an accumulation time of a second.
\end{abstract}
\maketitle

\section{Introduction}
Nuclear magnetic resonance (NMR) and magnetic resonance imaging (MRI) are attractive techniques to analyze properties of the nuclear spins and these techniques have a wide variety of the applications such as chemical analysis including determination of the protein structure, the study for molecular diffusion and biological imaging~\cite{ernst1988principlesofnmr,cavalli2007proteinstructuredetermination,abuayo1986NMRimaging,wuthrich1986NMRofPandNA}.
Typically, in these techniques, an oscillating magnetic field from the target nuclear spin ensemble is induced by irradiating radio frequency~(RF) pulses and the magnetic field is detected by a surrounding coil through inductive coupling.
There are many variations of the techniques to improve sensitivity and spatial resolution such as dynamic nuclear polarization~\cite{Overhauser53}, SQUID detected NMR~\cite{augustine1998SQUIDNMR}, MRFM~\cite{0957-4484-21-34-342001}, microslot waveguide NMR probe~\cite{Maguire2007} and an external high-Q resonator~\cite{martin2015EHQFresonator}.

Recently, a new approach to detect nuclear spins by using an electron spin of the nitrogen-vacancy~(NV) center in diamond has been demonstrated~\cite{kolkowitz2012nmrsensingSingleElectronSpin,mamin2013NanoscaleNMR,staudacher2013NMRon5nanometer}.
The NV center is used as an effective two-level system~(qubit) and has a long coherence time such as 2~milliseconds at a room temperature~\cite{balasubramanian2009ultralong,mizuochi2009coherence,bar-gill2013onesecondcoherence}.
It can be controlled by the microwave pulses and can be read out via the detection of photoluminescence from the NV center at the room temperature.
The nuclear spins with zero or almost zero polarization have Larmor precession to induce AC magnetic fields with random fluctuating amplitude and phase.
Such a randomized AC magnetic field can be detected by implementing a spin echo or dynamical decoupling on the NV centers~\cite{mamin2013NanoscaleNMR,staudacher2013NMRon5nanometer,muller2014NMRwithsinglespin}.
In these schemes, intervals of $\pi$ pulses are swept so that the resonance can be observed when the inverse of the intervals corresponds to twice the Larmor frequency of the nuclear spins.
Since the NV center has the long coherence time and the strong coupling strength due to the short distance between the NV center and nuclear spin, the sensitivity of such NMR is approaching a level of a single nuclear spin detection~\cite{muller2014NMRwithsinglespin}.
In the sensing approach using qubits, the sensitivity can be improved by entangling qubits~\cite{degan2017QuantumSensing} and the entanglement between NV centers has been extensively studied~\cite{Dolde2013EntanglementNVcenterExperiment, Yao2012EntanglementNVcenterTheory}.
However, since the NV center is coupled with the nuclear spins via a dipole-dipole interaction whose strength decreases by $1/r^3$, where $r$ denotes the distance between them, the NV center can only detect nuclear spins with a distance of tens of nanometers in the current technology.

In this paper, we propose an approach to detect nuclear spins by using a superconducting flux qubit~(FQ)~\cite{orlando1999firstFQtheory}.
The FQ is an artificial atom with a size of a few~$\mu$m.
The FQ has been considered as one of the promising systems to realize a quantum computer.
Extensive efforts have been devoted to improve performance of the FQ~\cite{Chiorescu2003CoherentOscillations,stern2014flux3Dcavity,yan2016FQrevisited,chiorescu2004coherent,plantenberg2007demonstration,lupacscu2004nondestructive,lupacscu2006high,paauw2009tuning} and multi-qubit entanglement has been realized~\cite{Lanting2014EntanglentInFluxQubit}.
It is possible to implement a single qubit rotation with high fidelity, and also we can read out the FQ by using a microwave resonator or a Josephson bifurcation amplifier where a readout visibility can reach more than $80\%$~\cite{you2007c-shunttheory,lupascu2007DelftJBA,stern2014flux3Dcavity,yan2016FQrevisited}.
The frequency of the FQ can be shifted by changing a magnetic
flux penetrating a qubit loop.
Therefore, we can measure magnetic fields by using the FQ \cite{bal2012ultrasensitive}.
There is inductive coupling between the FQ and an electron spin ~\cite{Marcos2010couplingNVtoFQ,twamley2010superconducting,zhu2011coherentcoupling,shiro2013quantummemorySQ,pramatsuzaki2015improving}.
The coupling strength is approximately scaled as $1/r$ as long as $r$ is comparable or smaller than the characteristic length of the FQ, where $r$ denotes the distance between the FQ and the spin.
Hence it is in principle possible to detect the spin far from the FQ~\cite{Marcos2010couplingNVtoFQ,twamley2010superconducting,matsuzaki2015improving}.
There are many potential applications by using this property such as a quantum memory~\cite{Marcos2010couplingNVtoFQ,twamley2010superconducting,zhu2011coherentcoupling,kubo2011hybrid,matsuzaki2012enhanced}
or magnetic field sensing~\cite{tanaka2015proposed,dooley2016hybrid}.
Although there are several types of researches to detect local electron spins using superconducting resonators~\cite{schuster2010high,kubo2010strong,kubo2012electron,bienfait2016reaching,probst2017inductive}, it is discussed that the FQ has a reasonable advantage to detect electron spins in a narrower region with high sensitivity~\cite{toida2017ESRwithFQ}.
Recently, by using the FQ as a detector of magnetization of electron spins, electron spin resonance~(ESR) was demonstrated, and hundreds of the electron spin with a volume of 50 femtoliters can be detected by a total accumulation time of a second~\cite{toida2017ESRwithFQ}.
These results show the excellent potential of the FQ to detect nuclear spins, and we theoretically investigate the performance of the FQ for NMR.

\begin{figure}[t]
  \begin{center}
\includegraphics[width=8cm]{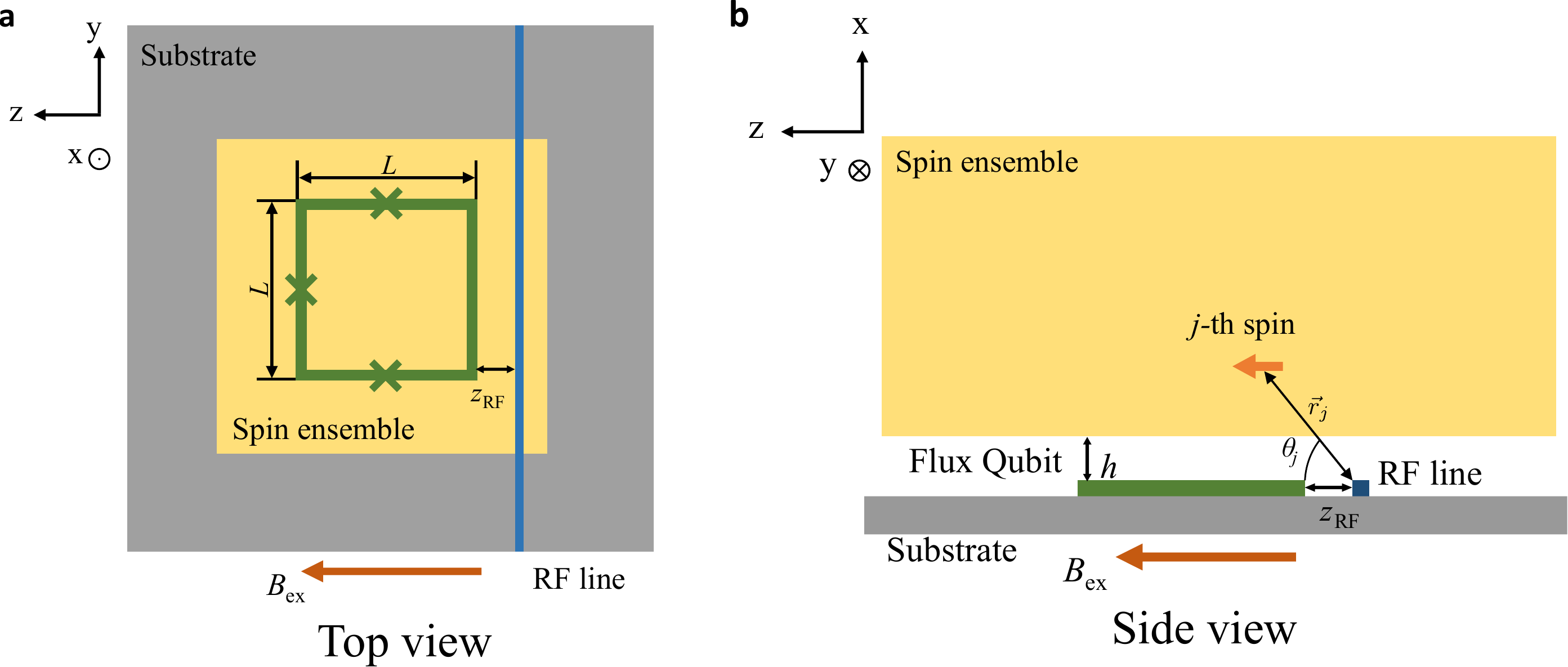}
\end{center}
\caption{
Schematic of our setup to implement NMR with a FQ from the top view~{\bf a} and the side view~{\bf b}.
A spin sample containing target nuclear spins is attached on the FQ.
A static magnetic field is applied along the z-direction to polarize the nuclear spins.
We will drive the nuclear spins by AC currents through the RF line until the nuclear spin system reaches a steady state.
The side of the FQ square is a length of $L$.
The distance between the spin sample and the FQ is $h$.
A line to apply RF pulses is located with a distance of $z_{\rm{RF}}$ from the FQ, and $B\u{ex}$ is the external magnetic field applied for polarizing the nuclear spins.
Since $B\u{ex}$ is applied in the z-direction, this does not affect the penetrating magnetic flux of the loop of the FQ.
We define the distance $r_j = |\vec{r_j}|$ between the $j$-th spin and the RF line and the angle $\theta_j$ between $\vec{r}    _j$ and z axis.
}
\label{fig:FQsetup}
\end{figure}

We consider two schemes for NMR with the FQ to analyze their performance.
In the first scheme, the FQ detects a DC magnetic field from the nuclear spins by using spatially inhomogeneous excitations of the nuclear spins.
This method has been used to realize the electron spin resonance with the FQ~\cite{toida2017ESRwithFQ}.
In this scheme, we use an on-chip RF line near the FQ for driving the nuclear spins.
The schematic of our setup is shown in Fig.~\ref{fig:FQsetup}.
The essential idea is that the partial excitation of the spins by the asymmetric driving induces a difference of the DC magnetic flux penetrating the loop of the FQ due to the driving.
In the second scheme, the FQ detects an AC magnetic field from the nuclear spins which are induced from the Larmor precession of the nuclear spins.
Here, we use a dynamical decoupling on the FQ to detect the AC magnetic fields.
This approach has been used to demonstrate NMR with the NV centers in diamond as previously discussed~\cite{mamin2013NanoscaleNMR}.

Our paper is organized as follows.
In Sec.~II, we review the standard general magnetic field sensing schemes with a qubit.
In Sec.~III, we describe NMR spectroscopy with an FQ using these two schemes.
In Sec.~IV, we show our numerical results for the minimum detectable density and the minimum detectable number of the nuclear spins.
In Sec.~V, we conclude our discussion.

\section{Magnetic field sensing with a qubit}
Here, we review standard sensing schemes to detect either DC or AC magnetic field with a qubit~\cite{degan2017QuantumSensing}.

\subsection{DC magnetic field sensing}
Suppose that a frequency of a qubit is shifted by magnetic fields so that the Hamiltonian in the rotating frame of the qubit frequency can be described by
\begin{align}
\hat{H}\u{DC}=\frac{\Delta \omega\u{DC}}{2}
 \hat{\sigma
 }_{z}
\end{align}
where $\hat{\sigma
}_{z}
=\ket{0}\bra{0}-\ket{1}\bra{1}$
is the Pauli $Z$ operator,
$\Delta \omega\u{DC} =\omega'\u{DC}-\omega\u{DC}$ denotes a detuning of the qubit, and $\omega\u{DC}$ ($\omega'\u{DC}$) is a frequency of the qubit without (with) an applied DC magnetic field.
$\omega'\u{DC}$ is a function of the applied magnetic fields.
Throughout this paper, we set $\hbar=1$.
The basic strategy for the sensing is to know the deviation of the frequency from the original one $\omega\u{DC}$.
First, prepare $\ket{+}=(\ket{0}+\ket{1})/\sqrt{2}$ state by applying a $\frac{\pi}{2}$ pulse to the qubit.
Second, let this state evolve by the Hamiltonian for a time $\tau $.
Finally, we readout the state by a projection operator described by $\hat{\mathcal{P}}_y=\frac{1+\hat{\sigma}_{y}}{2}$.
By repeating these processes within a total time $T\u{tot}$, we can obtain the average value of the projective measurements.
Since the expectation value of $\hat{\mathcal{P}}_y$ has a dependence on $\Delta \omega \u{DC}$, we can derive the value of $\Delta\omega\u{DC}$ and estimate DC magnetic field from the average of them.

\subsection{AC magnetic field sensing}
To detect AC magnetic fields, we can perform a dynamical decoupling on a qubit.
The Hamiltonian of the qubit with the applied AC magnetic field in a rotating frame is described as
\begin{align}
\hat{H}\u{AC}=\frac{\lambda\u{AC}}{2}\cos(\omega\u{AC} t)
 \hat{\sigma}\u{z},
\end{align}
where $\lambda\u{AC}$ is the amplitude of the change in the energy bias of the qubit due to the AC magnetic field and $\omega\u{AC}$ is the frequency of the AC magnetic field.
We can implement the dynamical decoupling
 on the qubit by using the following sequence.
First, prepare a $\ket{+}$ state by applying a $\frac{\pi}{2}$ pulse to the qubit.
Second, let this state evolve by the Hamiltonian for a time $\tau$ while we perform $\pi$ pulses with time $\tau$.
Finally, we read out the state by a projection operator described by $\hat{\mathcal{P}}_x=\frac{1+\hat{\sigma}_{x}}{2}$.
It is worth mentioning that the time interval of the $\pi$ pulses should be approximately set as $\tau \simeq 2\pi /\omega\u{AC} $ so that the qubit flip interval can synchronize with the AC magnetic field for the sensitive detection, where we assume the pulse lengths are much shorter than $\tau$.
Similar to the DC magnetic field sensing, by repeating these processes within the total time $T\u{tot}$, we can experimentally obtain the average value of the projective measurements.
Since the expectation value of $\hat{\mathcal{P}}_x$ has a dependence on the $\lambda_{\rm{AC}}$, we can estimate the amplitude of the AC magnetic fields from the average value.

\section{NMR sensing scheme with a Flux Qubit}
Here, we describe two sensing schemes to detect the NMR signal with the FQ.
The first scheme uses the DC magnetic field sensing and we call this a Ramsey measurement with asymmetric driving.
The other scheme uses a spin echo or a dynamical decoupling on the FQ to detect AC magnetic fields induced by the Larmor precession of the nuclear spins.
The schematic of our setup is shown in Fig.~\ref{fig:FQsetup}.
A spin sample containing nuclear spins is directly attached on the FQ.
The gyromagnetic ratio of the proton is the largest among typical nuclear spins.
This means that, as a proof of principle experiment of NMR using the FQ, it is suitable to use the protons as the target spins.
Therefore, throughout this paper, we consider the spin sample which includes the proton spins homogeneously.

\begin{figure}[t]
  \begin{center}
\includegraphics[width=8cm]{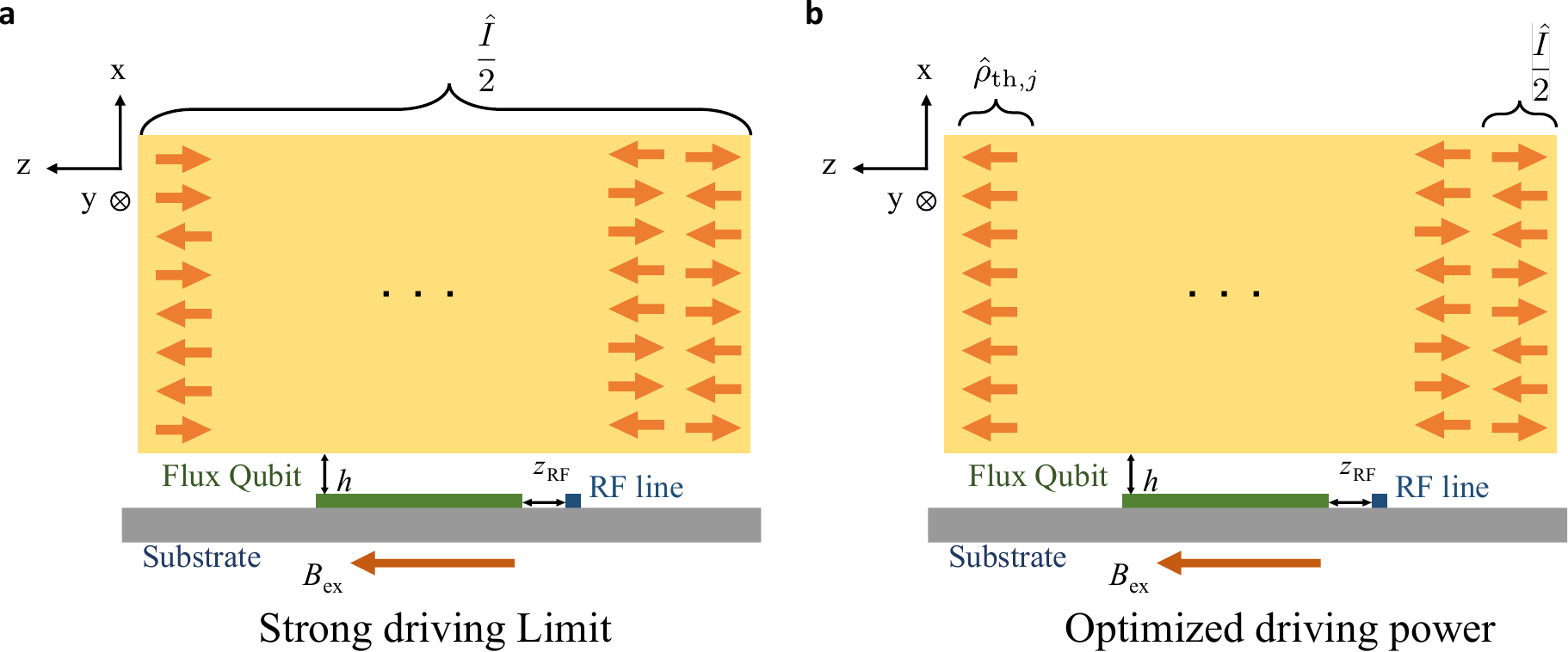}
\end{center}
\caption{
Schematic of the nuclear spins excited by asymmetric driving.
The arrows in the spin sample denote the nuclear spins which are polarized to parallel or anti-parallel to the $B\u{ex}$.
{\bf a}, In the strong driving limit, the nuclear spins are completely unpolarized, which is represented by  $\hat{I}/2$,  where the signals from the nuclear spins to the FQ are cancelled out.
{\bf b}, With optimized driving power, the polarization rate of the nuclear spins are spatially inhomogeneous so that the FQ can detect the change in DC magnetic fields penetrating the loop of the FQ.
The $\hat{\rho}_{\U{th},j}$ is the state of the $j$-th nuclear spin at a thermal state.
Although we consider a zero temperature case in this figure for simplicity, our idea can be also applied with a finite
 temperature case.
}
\label{fig:Asymmetricity}
\end{figure}

\subsection{NMR using Ramsey measurement with asymmetric driving}
We describe NMR using a Ramsey measurement with asymmetric driving.
The FQ detects a magnetic flux penetrating the qubit loop.
The magnetic flux is derived by integrating the x component of magnetic flux from spins.
Here, we consider the case that the size of the spin samples is large enough compared to the FQ.
When we drive the spins asymmetrically,  the total magnetic flux from the spins arises, and this generates changes in the FQ signals before and after the driving as shown in the Fig.~\ref{fig:Asymmetricity}.
On the other hand, if all spins become completely mixed states due to the strong driving, the total magnetic flux penetrating the qubit loop is cancelled out, and the FQ cannot obtain any signal change.

The Hamiltonian of our sensing system is written by using the Hamiltonian for the FQ, the spins, and the interaction between them as bellow,
\begin{align}
&\hat{H}=\hat{H}^{\U{(FQ)}}+\hat{H}^{\U{(int)}}+\hat{H}^{\U{(spin)}},\\
&\hat{H}^{\U{(FQ)}}=\frac{\epsilon}{2}\hat{\sigma^{\prime}}_{z}^{(\U{FQ})}+
\frac{\Delta}{2}\hat{\sigma^{\prime}}_{x}^{(\U{FQ})},\\
&\hat{H}^{\U{(int)}}=\frac{1}{2} \sum_{j=1}^{M} \gamma (\bm{B}_{j}^{(\U{FQ})} \cdot
\hat{\bm{\sigma}}_{j}^{\U{(spin)}})\hat{\sigma^{\prime}}_{z}^{\U{(FQ)}},\\
&\hat{H}^{\U{(spin)}}=\frac{1}{2}\sum_{j=1}^{M}\left(\omega_{j}\hat{\sigma}_{z,j}^{\U{(spin)}}+
4\lambda_{\U{RF},j}\hat{\sigma}_{x, j}^{\U{(spin)}}\cos \omega\u{RF}t\right).
\label{Hamiltonian}
\end{align}
Here $\epsilon=2I\u{p}(\Phi-\Phi\u{0}/2)$ is the frequency detuning,
$I\u{p}$ is the persistent current of the FQ,
$\Phi=B_{\U{ex}, \perp} L^2$ is the magnetic flux penetrating the FQ,
$\Phi\u{0}$ is the magnetic flux quanta,
$\hat{\sigma^{\prime}}_{z(x)}^{(\U{FQ})}$ is the Pauli Z (X) operator of the FQ,
$B_{\U{ex}, \perp}$, which is used for qubit control, is the x component of the external magnetic field,
$\Delta$ is the gap frequency of the flux qubit,
$\hat{\sigma}_{z,j}^{(\U{spin})}$ is the Pauli $Z$ operator for the $j$-th spin,
$\hat{\bm{\sigma}}_j^{(\U{spin})}=(\hat{\sigma}_{x,j}^{(\U{spin})},\hat{\sigma}_{y,j}^{(\U{spin})},\hat{\sigma}_{z,j}^{(\U{spin})})$ is the spin vector of the $j$-th spin,
$M$ is the total number of the spins,
$\gamma $ is the gyromagnetic ratio of the spins,
$\bm{B}_{j}^{\U{(FQ)}}$ is the magnetic field induced by the FQ at the $j$-th spin,
$\omega_{j}=\omega + \delta \omega_j=\gamma B\u{ex}^{(j)}$ is the Larmor frequency of the $j$-th spin,
$B\u{ex}^{(j)}=B\u{ex} +\delta B\u{ex}^{(j)}$ is the external magnetic field at the $j$-th spin,
$\omega=\gamma B\u{ex}$ is the average frequency of the spins,
$\delta\omega_j=\gamma \delta B\u{ex}^{(j)}$ is the frequency deviation of the $j$-th spin from the average,
$\delta B\u{ex}^{(j)}$ denotes randomized local magnetic field from the environment at the $j$-th spin,
$\lambda_{\U{RF},j}=\frac{\gamma \mu_0 I\u{RF}}{2\pi r_j}\cos \theta_j$ is the coupling strength of the $j$-th spin with the RF line,
$\mu\u{0}$ is the vacuum permittivity,
$\vec{r}_j$($r_j$) is the vector (distance) from the RF line to the spin,
$\theta_j$ is the elevation angle between the FQ surface and $\vec{r}_j$ (as shown in Fig.~\ref{fig:FQsetup}),
and $I\u{RF}$ is the current in the RF line.
We can diagonalize the flux qubit term by using $\frac{\omega\u{FQ}}{2} \hat{\sigma}^{\U{(FQ)}}_z
=\frac{\sqrt{\epsilon^2+\Delta^2}}{2}\hat{\sigma}^{\U{(FQ)}}_z
=\frac{\epsilon}{2}\hat{\sigma^{\prime}}_{z}^{(\U{FQ})}+
\frac{\Delta}{2}\hat{\sigma^{\prime}}_{x}^{(\U{FQ})}$ and
$\frac{\omega\u{FQ}}{2} \hat{\sigma}^{\U{(FQ)}}_x
=\frac{\Delta}{2}\hat{\sigma^{\prime}}_{z}^{(\U{FQ})}-
\frac{\epsilon}{2}\hat{\sigma^{\prime}}_{x}^{(\U{FQ})}$, as
\begin{align}
\hat{H}&=\frac{\omega\u{FQ}}{2}\hat{\sigma}_{z}^{(\U{FQ})}\nonumber \\
&+\frac{1}{2} \sum_{j=1}^{M} \gamma (\bm{B}_{j}^{(\U{FQ})} \cdot
\hat{\bm{\sigma}}_{j}^{\U{(spin)}})\left(\frac{\epsilon}{\omega\u{FQ}}\hat{\sigma}_{z}^{\U{(FQ)}}+\frac{\Delta}{\omega\u{FQ}}\hat{\sigma}_{x}^{\U{(FQ)}}\right)\nonumber \\
&+ \frac{1}{2}\sum_{j=1}^{M} \left(\omega_{j}\hat{\sigma}_{z,j}^{\U{(spin)}}+
4\lambda_{\U{RF},j}\hat{\sigma}_{x,j}^{\U{(spin)}}\cos \omega\u{RF}t\right).
\label{Hamiltonian}
\end{align}
This is the simplified Hamiltonian for the FQ and nuclear spins.
Next, we consider the Hamiltonian for a Ramsey measurement with asymmetric driving.
In a rotating frame for the FQ and the spins, which rotates at the frequency of $\omega^{\prime}$ and $\omega _j$,
this Hamiltonian becomes
\begin{align}
\hat{H}\u{AD}&\simeq\frac{\omega\u{FQ}-\omega^{\prime}}{2}\hat{\sigma}_{z}^{(\U{FQ})}\nonumber \\
&+ \frac{1}{2}\sum_{j=1}^{M} \left(\tilde{\gamma}(B_{z,j}^{(\U{FQ})}\hat{\sigma}_{z,j}^{\U{(spin)}})\hat{\sigma}_{z}^{\U{(FQ)}}
+2\lambda_{\U{RF},j}\hat{\sigma}_{x,j}^{\U{(spin)}}\right),
\end{align}
where $\tilde{\gamma}= \gamma\epsilon/\omega\u{FQ}$, we use the rotating-wave approximation for the FQ and the spins, while assuming that $\omega\u{RF}=\omega$ and
$1/\delta \omega_j$ is much larger than the time scale of this sequence.
The coupling strength between the FQ and the $j$-th spin $\tilde{\gamma}
B_{z,j}^{\U{(FQ)}}$ can be seen as the energy splitting of the FQ due to
the effective
DC magnetic field from the $j$-th spin $\gamma^{\prime}B_{z,j}^{\U{(spin)}}$.
Here $\gamma^{\prime}=\frac{d \omega\u{FQ}}{d B_{ \perp}}$
is the derivative of the qubit frequency with respect to the magnetic field $B_{ \perp}$ penetrating the loop of the FQ, and $B_{z,j}^{\U{(spin)}}$ denotes the DC magnetic field at the FQ induced by $j$-th spin.
When we consider a case without driving the nuclear spins ($\lambda_{\U{RF},j}=0$), we can simplify the Hamiltonian
as follows.
\begin{align}
  \hat{H}\u{AD} = \frac{1}{2} \left( \omega\u{FQ}-\omega^{\prime}+\gamma^{\prime}\sum_{j=1}^{M} B_{z,j}^{\U{(spin)}}\hat{\sigma}_{z,j}^{(\U{spin})} \right)\hat{\sigma}_{z}^{(\U{FQ})}.
\end{align}
Here, we assume that the nuclear spins  reach a thermal equilibrium state so that we can classically treat the nuclear spins.
By tracing out the freedom of the nuclear spin state,
only the magnetization from the nuclear spin state remains in the Hamiltonian to affect the dynamics of the FQ.
Especially, we define $\avg{\hat{\sigma}^{\U{(spin)}}_{z,j}}\u{th}$  as an expectation value of Pauli $Z$ operator for the $j$-th spin in the case of the thermalized nuclear-spin state without the RF driving.
On the other hand, we define $\avg{\hat{\sigma}^{\U{(spin)}}_{z, j}}\u{ st}$ as an expectation value of Pauli $Z$ operator for the $j$-th spin when the nuclear-spin state is in steady state by the RF driving.

\begin{figure}[t]
\includegraphics[width=8cm, clip]{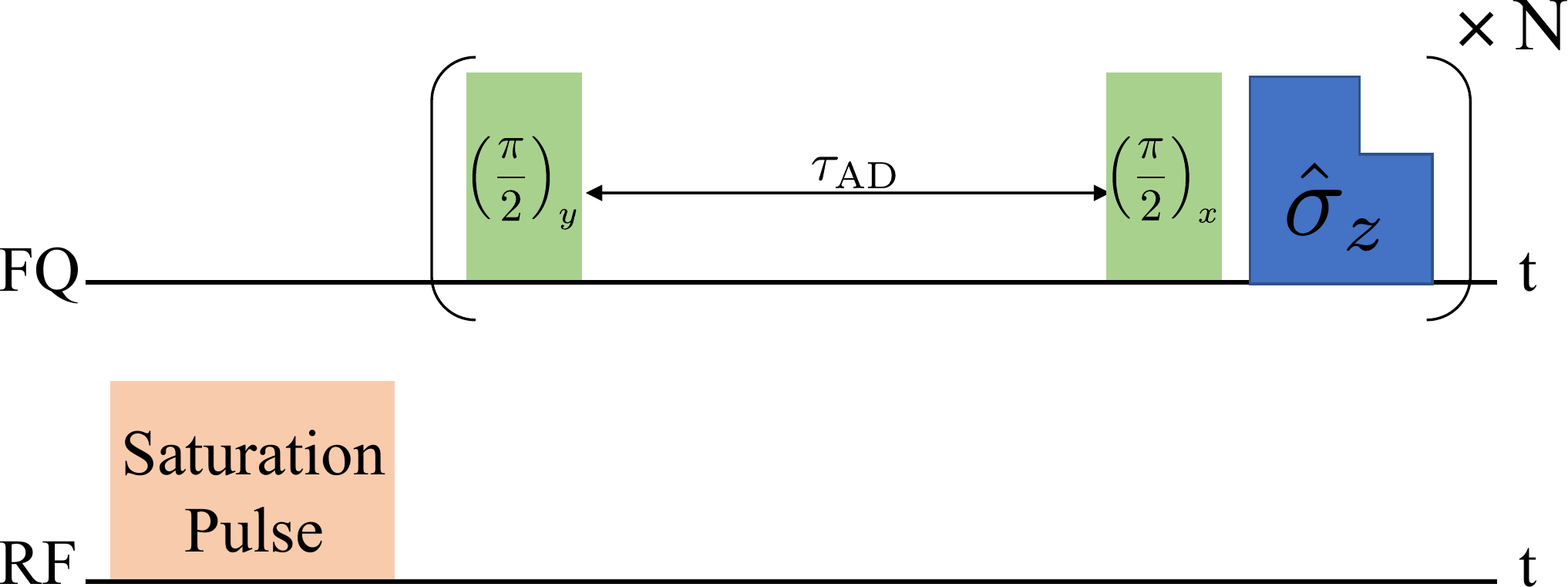}
\caption{The pulse sequence for a Ramsey measurement with asymmetric driving.
After the nuclear spins are excited by the RF pulse, the Ramsey interference
measurements performed N times.
$\hat{\sigma}_z$ means the measurement in the z-direction.
}
\label{fig:ADsequence}
\end{figure}

We can use a pulse sequence of the standard DC magnetic field sensing
for the detection of the nuclear spins as shown in Fig.~\ref{fig:ADsequence}.
In our scheme, the difference of the spin polarization between before and after the RF driving induces an effective DC magnetic fields to the FQ.
By setting $\omega^{\prime}=\omega\u{FQ}+\sum_{j=1}^{M}\gamma^{\prime} B_{z,j}^{\U{(spin)}}\avg{\hat{\sigma}_{z,j}^{{\rm (spin)}}}_{\rm{th}}$, the detuning caused by the RF driving is defined as
\begin{align}
\Delta \omega\u{FQ} &= \gamma^{\prime}\sum_{j=1}^{M} B^{\U{(spin)}}_{z,j}\left(\avg{\hat{\sigma}_{z,j}^{\U{(spin)}}}\u{st}-
\avg{\hat{\sigma}_{z,j}^{\U{(spin)}}}\u{th}\right),
\label{eq:difomega}
\end{align}
and the Hamiltonian for the Ramsey measurement becomes
\begin{align}
  \hat{H}\u{AD}^{\prime} = \frac{1}{2} \Delta \omega\u{FQ} \hat{\sigma}_{z}^{(\U{FQ})}.
\end{align}

In the end, the signal $P\u{AD}$ is calculated as
\begin{align}
  P\u{AD} &=\mathrm{Tr}[\mathrm{e}^{-i\hat{H}\u{AD}^{\prime}\tau\u{AD}} \ket{+}\bra{+}^{\U{(FQ)}} \mathrm{e}^{i\hat{H}\u{AD}^{\prime}\tau\u{AD}} \hat{\mathcal{P}}_y^{\U{(FQ)}}] \nonumber \\
  &=\frac{1}{2}+\frac{1}{2} \sin \Delta \omega _{\rm{FQ}} \tau\u{AD},
  \nonumber \\
  &\simeq \frac{1}{2}+\frac{1}{2} \Delta \omega _{\rm{FQ}} \tau\u{AD},
  \label{eq:signalAD}
\end{align}
where $\ket{+}\bra{+}^{\U{(FQ)}} = \ket{+}^{\U{(FQ)(FQ)}}\bra{+}$ and we assume that $\Delta \omega _{\rm{FQ}} \tau\u{AD}\ll 1$.

To maximize the detuning
$\Delta \omega\u{FQ}$, we
optimize the position of the RF line and the Rabi frequency.
For this purpose, we need to calculate $\avg{\hat{\sigma}^{\U{(spin)}}_{z,j}}\u{st}$ and $\avg{\hat{\sigma}^{\U{(spin)}}_{z,j}}\u{th}$.
The density matrix for the $j$-th spin $\hat{\rho}_j$ at the thermal state is calculated by using Boltzmann distribution as $\hat{\rho}_{\U{th},j}={\rm{exp}}\left[-\frac{\omega _j\hat{\sigma }_{z,j}^{(\rm{spin})}}{2k_B T}\right]/Z$ and $Z={\rm{Tr}}\left[{\rm{exp}}\left[-\frac{\omega _j\hat{\sigma}_{z,j}^{(\rm{spin})}}{2k_B T}\right]\right]$,
where $k_B$ is the Boltzmann constant and $T$ is the temperature.
We can get $\avg{\hat{\sigma}^{\U{(spin)}}_{z,j}}\u{th}={\rm{Tr}}\left[\hat{\sigma}_{z,j}^{\U{(spin)}}\hat{\rho}_{\U{th},j}\right]\simeq \frac{\omega}{k_{B}T}$, where we assume $\delta\omega_j\ll\omega$.

We will solve the Lindblad master equation of the $j$-th spin for calculating $\avg{\hat{\sigma}^{\U{(spin)}}_{z,j}}\u{st}$.
It is worth mentioning that, while we drive the nuclear spins by the RF pulses, the FQ is in a ground state.
In this case, we can trace out the FQ term from the Hamiltonian by taking $\hat{\sigma
}_z^{(\rm{FQ})}=-1$.
Then, the master equation is given as
\begin{align}
  \frac{d \hat{\rho_j}}{dt}= -i[\hat{H}_j,\hat{\rho}_j] + \hat{\mathcal{L}}_j\hat{\rho}_j,
  \label{eq:lindbladME}
\end{align}
where $\hat{H}_j$ is the Hamiltonian for the $j$-th spin and $\hat{\mathcal{L}}_j$ is the Lindblad superoperator for the $j$-th spin.
With RF driving, the Hamiltonian for the $j$-th spin is
\begin{align}
  \hat{H}_j= \frac{1}{2} \left(\omega +\delta\omega_j + \tilde{\gamma} B_{z,j}^{\U{(FQ})} \right) \hat{\sigma}_{z,j}^{{\rm (spin)}}\nonumber \\
  +2\lambda_{\U{RF},j}\hat{\sigma}_{x,j}^{\U{(spin)}}\cos \omega\u{RF}t.
\end{align}
In a rotating frame for the nuclear spin, the Hamiltonian is described as
\begin{align}
  \hat{H}_{j,\U{rot}}= \frac{1}{2} \delta \omega_j \hat{\sigma}_{z,j}^{{\rm (spin)}}+ \lambda_{\U{RF},j}\hat{\sigma}_{x,j}^{\U{(spin)}},
\end{align}
where we assume that $\omega\u{RF}=\omega$ and $\delta\omega_j\gg\tilde{\gamma} B_{z,j}^{\U{(FQ})}$, and we use the rotating-wave approximation.
The superoperator $\hat{\mathcal{L}}_j$ is described as
\begin{align}
  \hat{\mathcal{L}_j}\hat{\rho}_j= &-\frac{\Gamma}{2}(1-s)[\hat{\sigma}_{+,j} \hat{\sigma}_{-,j} \hat{\rho}_j + \hat{\rho}_j \hat{\sigma}_{+,j} \hat{\sigma}_{-,j}
  -2 \hat{\sigma}_{-,j} \hat{\rho}_j \hat{\sigma}_{+,j}] \nonumber \\
  &-\frac{\Gamma}{2}s[\hat{\sigma}_{-,j} \hat{\sigma}_{+,j} \hat{\rho}_j + \hat{\rho}_j \hat{\sigma}_{-,j} \hat{\sigma}_{+,j}
  -2 \hat{\sigma}_{+,j} \hat{\rho}_j \hat{\sigma}_{-,j}],
\end{align}
where $\Gamma$ is the longitudinal relaxation rate, $\hat{\sigma}_{+,j} =\hat{\sigma}_{-,j}^{\dagger}= \ket{1}_{jj}\bra{0}$ is the raising operator
and $s=\frac{1}{2}+\frac{1}{2}\avg{\hat{\sigma}^{\U{(spin)}}_{z,j}}\u{th}$ denotes a probability that the spin is excited at the thermal equilibrium state.
\begin{figure}[t]
\includegraphics[width=8cm, clip]{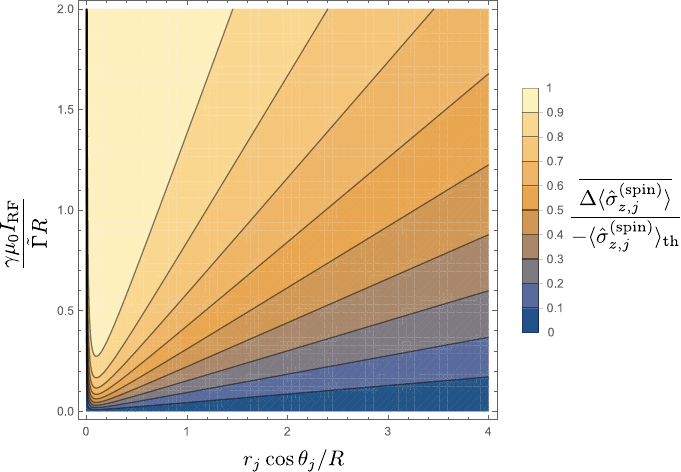}
\caption{The ratio of the polarization with RF driving
 $\overline{\Delta\avg{\hat{\sigma}^{\U{(spin)}}_{z,j}}}$ to the polarization
 of thermal state $\avg{\hat{\sigma}^{\U{(spin)}}_{z,j}}\u{th}$.
The vertical axis denotes $\gamma \mu_0 I\u{RF}/\tilde{\Gamma}R$ (normalized current strength of the RF pulse), while the horizontal axis denotes $r_j \cos\theta_j/R$ where $r_j \cos\theta_j$ denotes z component of the position vector $\vec{r}_j$  from RF line to the $j$-th spin (as shown in Fig. \ref{fig:FQsetup}).
In this calculation, we set the reference distance $R=1~\mu$m, temperature $T=20$~mK, the magnetic field $B\u{ex}=5$~mT and the x components of the $j$-th spin position is $0.1~\mu$m.
At $r_j \cos\theta_j =0$, since $\lambda_{\U{RF},j}=0$, $\overline{\Delta\avg{\hat{\sigma}^{\U{(spin)}}_{z,j}}}$ becomes zero.
}
\label{fig:$j$-th}
\end{figure}
\begin{figure}[t]
\includegraphics[width=8cm, clip]{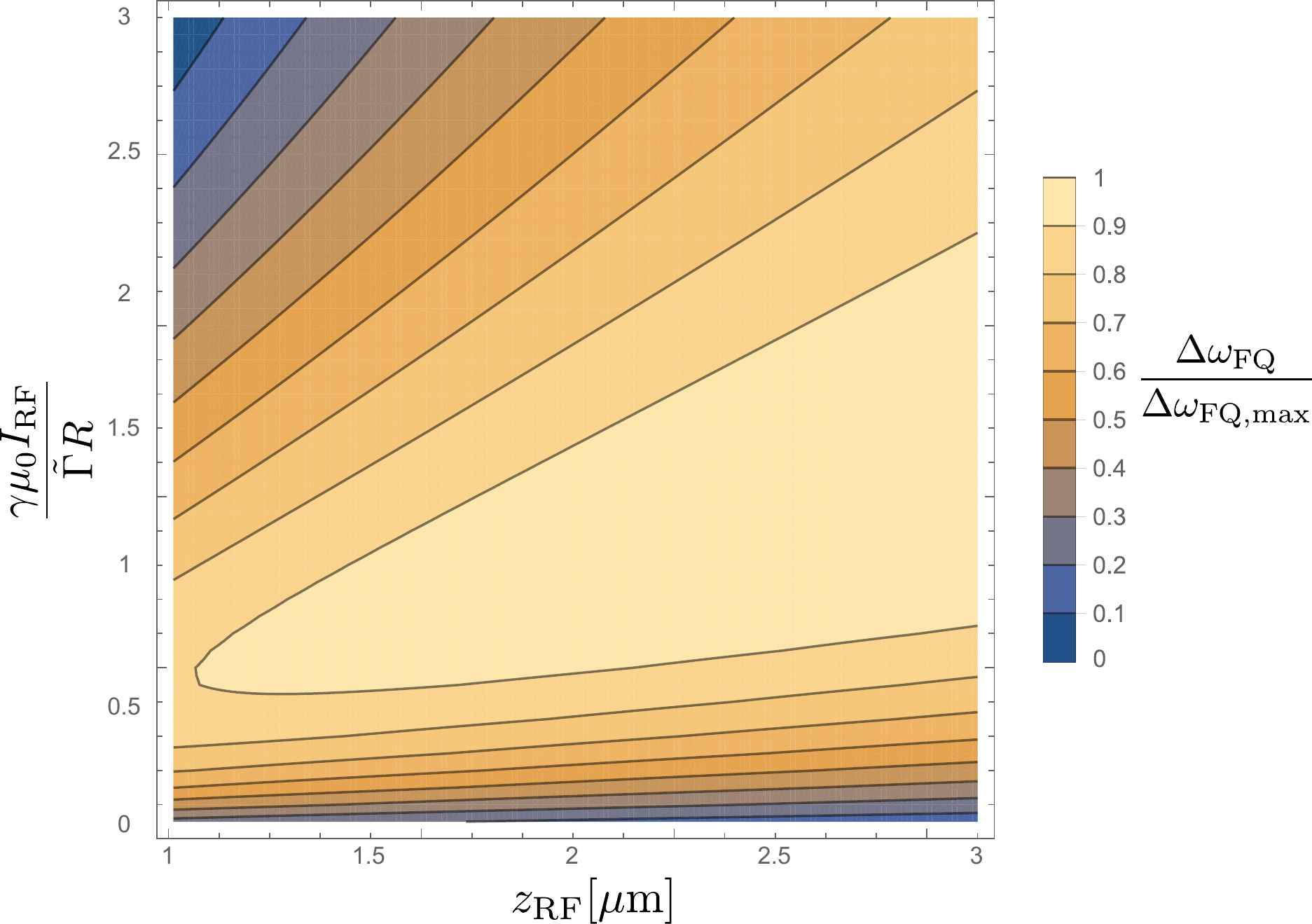}
\caption{The signal of the FQ as a function of $z\u{RF}$ and the $\gamma \mu_0 I\u{RF}/\tilde{\Gamma}R$.
$L$ and $h$ are 2~$\mu$m and 0.1~$\mu$m, respectively.
We use the same parameters as that used in the Fig.~4 except the parameters mentioned above.
The size of the spin sample is $2.5L\times2.5L\times2L$.
$\Delta \omega\u{FQ,max}$ is the maximum value of $\Delta\omega\u{FQ}$ in this plot region.}
\label{fig:RFline}
\end{figure}

For a given frequency deviation $\delta \omega _j$, we solve the master equation~(\ref{eq:lindbladME}) for the steady state, and obtain the polarization difference between the thermal and saturated state
\begin{align}
\Delta\avg{\hat{\sigma}_{z,j}^{\U{(spin)}}} &=\avg{\hat{\sigma}_{z,j}^{\U{(spin)}}}\u{st}-\avg{\hat{\sigma}_{z,j}^{\U{(spin)}}}\u{th}\nonumber \\
&=-\avg{\hat{\sigma}_{z,j}^{\U{(spin)}}}\u{th} \frac{8 \lambda_{\U{RF},j}^2}{\Gamma ^2 +8 \lambda_{\U{RF},j}^2 + 4 \delta \omega_j^2}.
\end{align}
However, in the real systems, the nuclear spins are affected by a low-frequency magnetic field noise $\delta B\u{ex}^{(j)}$ from the environment.
To take into account this effect, we consider an ensemble average of the frequency deviation with a Gaussian weight as follows.
\begin{align}
  \overline{\Delta\avg{\hat{\sigma}_{z,j}^{\U{(spin)}}}}
  &= -\avg{\hat{\sigma}_{z,j}^{\U{(spin)}}}\u{th}
  \int_{-\infty}^{\infty} \frac{1}{\tilde{\Gamma} \sqrt{\pi}}\mathrm{e}^{-\frac{\delta\omega^{2}_{j}}{\tilde{\Gamma}^{2}}}\nonumber \\
  &\qquad \times \frac{8 \lambda_{\U{RF},j}^2}{\Gamma ^2 +8 \lambda_{\U{RF},j}^2 + 4 \delta \omega_j^2} d (\delta\omega_j) \nonumber \\
  &=-\avg{\hat{\sigma}_{z,j}^{\U{(spin)}}}\u{th}
  \frac{4 \lambda_{\U{RF},j}^{2}\sqrt{\pi}\mathrm{e}^{\frac{\Gamma ^2+8\lambda_{\U{RF},j}^2}{4\tilde{\Gamma}^{ 2}}}}
  {\tilde{\Gamma} \sqrt{\Gamma^2+8\lambda_{\U{RF},j}^{2}}}\nonumber \\
  & \qquad \times \mathrm{Erfc}\left[\frac{\sqrt{\Gamma^2 +8 \lambda_{\U{RF},j}^{2}}}{2\tilde{\Gamma}}\right]
\end{align}
where $\tilde{\Gamma}$ is linewidth of the frequency $\delta\omega_j$ due to the environmental magnetic field noise and $\U{Erfc}[\cdot]$ is a complementary error function.
Since the energy relaxation is typically much weaker than the low-frequency magnetic field noise, we assume that $\Gamma \ll \tilde{\Gamma}$ throughout this paper~\cite{Angerere2017ultralongrelaxation,rangga2018phonon-bottleneck,amususs2011cavitycoupledspin}
and we set $\Gamma= 10^{-3}\times\tilde{\Gamma}$ in the calculation.
(It is worth mentioning that our results are not significantly changed for any value of $\Gamma $ as long as the condition of $\Gamma \leq 10^{-3}\times \tilde{\Gamma}$ is satisfied, which we numerically confirmed.)
We need a position dependence of $\overline{\avg{\hat{\sigma}_{z,j}^{\U{(spin)}}}}\u{st}$ to evaluate the effect of the spatially inhomogeneous excitation of the nuclear spins after the RF driving.
To illustrate such an asymmetric excitation, the density plot of normalized polarization $\overline{\Delta\avg{\hat{\sigma}_{z,j}^{\U{(spin)}}}}/(-\avg{\hat{\sigma}_{z,j}^{\U{(spin)}}}\u{th})$  is shown in Fig.~\ref{fig:$j$-th}.
As the nuclear spins are located closer to the RF line, the excitation ratio after the driving becomes larger so that the spin excitation ratio can be spatially inhomogeneous in our setup.

To optimize the position of the RF line and the current of the RF pulse, we plot $\Delta \omega _{\u{FQ}}/\Delta \omega _{\u{FQ,max}}$ in the Eq.~(\ref{eq:difomega}) as a function of $z\u{RF}$ and the $\gamma \mu_0 I\u{RF}/\tilde{\Gamma}R$ in Fig.~\ref{fig:RFline}.
This shows that $\Delta\omega\u{FQ}/\Delta\omega\u{FQ,max}$  is optimized when $\gamma \mu_0 I\u{RF}/\tilde{\Gamma} R\simeq 1$ is satisfied for $1$~$\mu$m $<z\u{RF}<3$~$\mu$m.
In the actual experiment, the current in the RF line can be as large as a few mA.
This means that, as long as $\tilde{\Gamma}<10^5$~s$^{-1}$, we can optimize the signal by controlling the $I\u{RF}$.
Therefore, throughout this paper, we fix the value of $z\u{RF}$, because we can obtain almost the same optimal signal by choosing $I\u{RF}$ for a given $z\u{RF}$ as shown in the Fig.~\ref{fig:RFline}.
Therefore, in the calculation section (Section~\ref{sec:CalculationSection}), we fix the value of $z\u{RF}$, because we can obtain almost the same optimal signal by choosing $I\u{RF}$ for a given $z\u{RF}$ as shown in the Fig.~\ref{fig:RFline}.

For more realistic estimation, we consider the effect of the dephasing
of the FQ and an imperfect readout.
We adopt a dephasing channel of the FQ such as
$\hat{\mathcal{E}}(\hat{\rho} )= p\hat{\rho} + (1-p) \hat{\sigma }_z^{(\rm{FQ})}\hat{\rho}\hat{\sigma }_z^{(\rm{FQ})}$
for a density matrix of the FQ $\hat{\rho}$, where
$p=\frac{1}{2}+\frac{1}{2}e^{-\Gamma\u{AD}^{\U{(FQ)}}\tau\u{AD}}$ denotes a probability to induce the dephasing during the interaction time $\tau\u{AD}$, $\Gamma\u{AD}^{\U{(FQ)}}=1/T_2^{\ast}$ denotes the dephasing rate of the FQ and $T_2^{\ast}$ is the dephasing time for a Ramsey measurement.
The qubit state before the readout step can be described as
\begin{align}
    \hat{\rho}\u{AD} &=\frac{1+\mathrm{e}^{-\Gamma\u{AD}^{\U{(FQ)}} \tau\u{AD}}}{2}\left(\mathrm{e}^{-i\hat{H}\u{AD}^{\prime}\tau\u{AD}} \ket{+}\bra{+}^{\U{(FQ)}} \mathrm{e}^{i\hat{H}\u{AD}^{\prime}\tau\u{AD}}\right) \nonumber \\
    &+ \frac{1-\mathrm{e}^{-\Gamma\u{AD}^{\U{(FQ)}} \tau\u{AD}}}{2} \nonumber \\
    &\times \left(\hat{\sigma}_z^{\U{(FQ)}} \mathrm{e}^{-i\hat{H}\u{AD}^{\prime}\tau\u{AD}} \ket{+}\bra{+}^{\U{(FQ)}} \mathrm{e}^{i\hat{H}\u{AD}^{\prime}\tau\u{AD}} \hat{\sigma}_z^{\U{(FQ)}}\right).
\end{align}
After the readout by $\hat{\mathcal{P}}_y$ we can get the signal as
\begin{align}
  \overline{P'}\u{AD} \simeq \frac{1}{2}+\frac{1}{2}\mathrm{e}^{-\Gamma\u{AD}^{\U{(FQ)}} \tau\u{AD}} \overline{\Delta \omega}\u{FQ} \tau\u{AD}, \\
  \overline{\Delta \omega}\u{FQ}=\gamma^{\prime}\sum_{j=1}^{M} B^{\U{(spin)}}_{z,j}\overline{\Delta\avg{\hat{\sigma}_{z,j}^{\U{(spin)}}}}.\nonumber
\end{align}

Suppose a perfect measurement apparatus~(MA) was available, the MA would provide us with a specific detection signal (such as a large electrical
current) if and only if the state of the FQ is
 $\ket{e}^{\U{(FQ)}}$ while the MA would not generate such detection signal with a state of $\ket{g}^{\U{(FQ)}}$, where $\ket{e(g)}^{\U{(FQ)}}$ denotes that the state of the FQ is the excited (ground) state.
However, the measurement apparatus is imperfect in the actual experiment, and the measurement results may not correspond to the actual state of the FQ.
To include such an imperfection, we adopt a model that the FQ is depolarized due to the interaction with the MA by the following error channel
\begin{align}
 \hat{\mathcal{E}}(\hat{\rho}\u{AD})
   =(1-\eta)\hat{\rho}\u{AD}+\eta \frac{\hat{I}}{2},
\end{align}
where $\eta$ is the depolarization ratio. We assume that a projective measurement can be implemented only after the FQ is affected by this error channel.
In this case, the signal can be described as
\begin{align}
  \overline{\tilde{P'}}\u{AD}&\simeq
  \frac{1}{2}+\frac{1}{2}(1-\eta) \mathrm{e}^{-\Gamma\u{AD}^{\U{(FQ)}}
  \tau\u{AD}}\overline{\Delta \omega}{\u{FQ}} \tau\u{AD}. \nonumber
\end{align}

In order to quantify the accuracy of the measurement process, we define a probability that the imperfect MA shows the detection signal (that is expected to occur when the FQ is excited) as $p(\U{detect})$.
Especially, we consider conditional probabilities such as $p(\U{detect}|\ket{e}^{\U{(FQ)}})$  [$p(\U{detect}|\ket{g}^{\U{(FQ)}})$] to observe the MA detection signal when the FQ state is prepared in $\ket{e}^{\U{(FQ)}}$  [$\ket{g}^{\U{(FQ)}}$].
By using our error model, we can calculate these as $p(\U{detect}|\ket{e}^{\U{(FQ)}})=1-\frac{\eta}{2}$ and $p(\U{detect}|\ket{g}^{\U{(FQ)}})=\frac{\eta}{2}$.
The so-called visibility $V$ is defined as $V=p(\U{detect}|\ket{e}^{\U{(FQ)}})-p(\U{detect}|\ket{g}^{\U{(FQ)}})$.
In our model, the visibility is described as $V=1-\eta$.
From this relationship, the signal of the FQ can be described as
\begin{align}
  \overline{\tilde{P'}}\u{AD}\simeq
  \frac{1}{2}+\frac{1}{2}V\mathrm{e}^{-\Gamma\u{AD}^{\U{(FQ)}} \tau\u{AD}}\overline{\Delta \omega}\u{FQ} \tau\u{AD}.
\end{align}

Next, we consider the optimization of the interaction time $\tau\u{AD}$.
In our scheme, we measure DC magnetic fields from the
nuclear spins. According to the standard prescription of the quantum
metrology~\cite{degan2017QuantumSensing}, we will consider the uncertainty of the estimation of the target fields as follows
\begin{align}
  \delta B_{\rm{DC}} &= \nonumber
 \frac{\sqrt{\overline{\tilde{P'}}\u{AD}(1-\overline{\tilde{P'}}\u{AD})}}
 {\left|\frac{d\overline{\tilde{P'}}\u{AD}}{dB\u{DC}}\right|\sqrt{N}}
\simeq \frac{\mathrm{e}^{\Gamma\u{AD}^{\U{(FQ)}} \tau\u{AD}}}{V\gamma^{\prime}\tau\u{AD}\sqrt{N}},
\end{align}
where
$B_{\rm{ DC}}=\sum_{j=1}^{M} B_{z,j}^{\U{(spin)}} \overline{\Delta\avg{\hat{\sigma}_{z,j}^{\U{(spin)}}}}$
is the effective DC magnetic field from the nuclear spins,
$N=T\u{tot}/T\u{rep}$ is the number of repetitions, and $T\u{rep}$ is the time required for a single measurement.
The interaction time $\tau\u{AD}$ to minimize this uncertainty is $\tau\u{AD}^{\U{opt}}=1/\Gamma\u{AD}^{(\U{FQ})}=T_2^{\ast}$.

\subsection{NMR using dynamical decoupling}
\begin{figure}[t]
\includegraphics[width=9cm, clip]{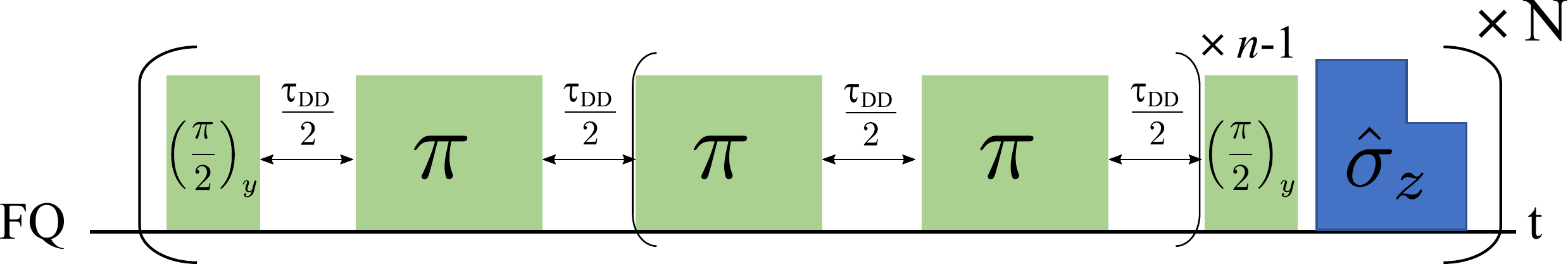}
\caption{The pulse sequence for dynamical decoupling using $n$ $\pi$ pulses.
}
\label{fig:PulsesequenceSE}
\end{figure}

We describe the NMR by using the dynamical decoupling on the FQ.
We adopt the same Hamiltonian as the Eq.(\ref{Hamiltonian}).
We detect Larmor precession of the nuclear spins, which induces AC magnetic fields.
We can use a pulse sequence shown in Fig.~\ref{fig:PulsesequenceSE} with $2n\!-\!1$ $\pi$ pulses.
This technique has been used to detect nuclear spins by using a single NV center~\cite{mamin2013NanoscaleNMR}.
It is worth mentioning that neither RF driving ($\lambda _{\rm{RF}}=0$) nor polarization of the nuclear spins is required for this detection.
For simplicity, we consider the case of single nuclear spin coupled with the FQ.
In a rotating frame for the FQ which rotates at the frequency of $\omega\u{FQ}$, the Hamiltonian in the Eq.(\ref{Hamiltonian}) for the FQ and the $j$-th spin becomes
\begin{align}
\hat{H}_j&\simeq \frac{1}{2}\left( \tilde{\gamma} B_{z, j}^{(\U{FQ})} \hat{\sigma}_{z}^{\U{(FQ)}}
+\omega_j\right) \hat{\sigma}_{z,j}^{(\U{spin})}\nonumber \\
&+\frac{\tilde{\gamma}}{2} B_{\perp, j}^{(\U{FQ})} \hat{\sigma}_{\perp, j}^{(\U{spin})}\hat{\sigma}_{z}^{\U{(FQ)}},\nonumber \\
\end{align}
 where $B_{\perp, j}^{(\U{FQ})}=\sqrt{\left(B_{x, j}^{(\U{FQ})}\right)^2+\left(B_{y, j}^{(\U{FQ})}\right)^2}$ is the amplitude of the magnetic field in an x-y plane and $\hat{\sigma}_{\perp,j}^{(\U{spin})}= B_{x,j}^{(\U{FQ})}/B_{\perp,j}^{(\U{FQ})} \hat{\sigma}_{x,j}^{(\U{spin})}+B_{y,j}^{(\U{FQ})}/B_{\perp,j}^{(\U{FQ},j)} \hat{\sigma}_{y,j}^{(\U{spin})}$.
 In a rotating frame for the spin, the last term can be regarded as a coupling of the AC magnetic field from the nuclear spins and the magnetic field from the FQ~\cite{ajoy2016DCmagnetmetry}.
Similar to the case of the Ramsey measurement with asymmetric driving,
we use a relationship of
$\tilde{\gamma}B_{\perp, j}^{(\U{FQ})}=\gamma^{\prime}B_{\perp,
j}^{(\U{spin})}$, and rewrite the Hamiltonian
where $B_{\perp, j}^{(\U{spin})}$ is the AC magnetic field effect from nuclear spins.
For $\omega_j \gg \tilde{\gamma} B^{\U{(FQ)}}_{z,j} $, the Hamiltonian is rewritten as
\begin{align}
\hat{H}_{\U{DD},j}&=\frac{\omega_j}{2} \hat{\sigma}_{z,j}^{(\U{spin})}
+\frac{\gamma^{\prime}}{2} B_{\perp,j}^{(\U{spin})} \hat{\sigma}_{\perp,j}^{\U{(spin)}}
\hat{\sigma}_{z}^{\U{(FQ)}} \nonumber \\
&= \ket{0}\bra{0}^{(\U{FQ})} \otimes \hat{H}_{0,j}^{(\U{spin})} + \ket{1}\bra{1}^{(\U{FQ})}\otimes \hat{H}_{1,j}^{(\U{spin})},\nonumber \\
\label{eq:SEHamiltonnian}
\end{align}
where $\hat{H}_{0,j}^{(\U{spin})} =\frac{\omega_j}{2}\hat{\sigma}_{z,j}^{\U{(spin)}}
+\frac{\gamma^{\prime}}{2} B_{\perp,j}^{(\U{spin})} \hat{\sigma}_{\perp,j}^{\U{(spin)}}$
and $\hat{H}_{1,j}^{(\U{spin})} =\frac{\omega_j}{2}\hat{\sigma}_{z,j}^{\U{(spin)}}
-\frac{\gamma^{\prime}}{2} B_{\perp,j}^{(\U{spin})} \hat{\sigma}_{\perp,j}^{\U{(spin)}}$.
We prepare an initial state of $\ket{\psi_{\pm, j}(0)}=\ket{+}^{\U{(FQ)}}\otimes\ket{\pm }_j^{(\U{spin})}$.
In this section, we consider a case of $n\!=\!1$, which is called a spin echo.
Let this evolve by the Hamiltonian for a time $\tau\u{DD}/2$, and we obtain
\begin{align}
  \left| \psi_{\pm,j}\left(\frac{\tau\u{DD}}{2}\right) \right\rangle
  =\frac{1}{\sqrt{2}}\ket{0}^{\U{(FQ)}} \otimes \mathrm{e}^{-i\hat{H}^{\rm{(spin)}}_{0,j} \frac{\tau\u{DD}}{2}}
   \ket{\pm}^{\U{(spin)}}_j \nonumber \\
  +\frac{1}{\sqrt{2}}\ket{1}^{\U{(FQ)}} \otimes \mathrm{e}^{-i\hat{H}^{\rm{(spin)}}_{1,j} \frac{\tau\u{DD}}{2}} \ket{\pm }^{\U{(spin)}}_j.
\end{align}
After performing a $\pi$ pulse on the FQ, let the state evolve for a time
$\tau\u{DD}/2$, and we obtain
\begin{align}
  \ket{\psi_{\pm,j}(\tau\u{DD})}=\frac{1}{\sqrt{2}}\ket{1}^{\U{(FQ)}} \otimes \hat{U}_a
   \ket{\pm }^{\U{(spin)}}_j \\ \nonumber
  +\frac{1}{\sqrt{2}}\ket{0}^{\U{(FQ)}} \otimes \hat{U}_b \ket{\pm }^{\U{(spin)}}_j.
\end{align}
where $\hat{U}_{a,j}= \mathrm{e}^{-i\hat{H}_{1,j}^{\U{(spin)}} \frac{\tau\u{DD}}{2}} \mathrm{e}^{-i\hat{H}_{0,j}^{\U{(spin)}} \frac{\tau\u{DD}}{2}}$ and
$\hat{U}_{b,j}= \mathrm{e}^{-i\hat{H}_{0,j}^{\U{(spin)}} \frac{\tau\u{DD}}{2}} \mathrm{e}^{-i\hat{H}_{1,j}^{\U{(spin)}} \frac{\tau\u{DD}}{2}}$.
By reading out the state of the FQ with a projection operator $\hat{\mathcal{P}}_x = \frac{1+\hat{\sigma}_x}{2}$,
we have
\begin{align}
  P_{\pm}&=\mathrm{Tr}\left[\hat{\mathcal{P}}_x \ket{\psi_{\pm,j   }(\tau\u{DD})}\bra{\psi_{\pm,j}(\tau\u{DD})}\right]\nonumber \\
  &=\frac{1}{2}+\frac{1}{4}\ _{\quad j}^{\U{(spin)}}\bra{\pm} (\hat{U}_a^{\dagger}\hat{U}_b+\hat{U}_b^{\dagger}\hat{U}_a)\ket{\pm}_j^{\U{(spin)}}.
\end{align}
So, the signal will be calculated as
\begin{align}
  P_{\pm}\simeq 1-\left( \cos \frac{\omega_j \tau\u{DD}}{2}-1\right)^2 \frac{ \left(\gamma^{\prime} B_{\perp,j}^{(\U{spin})} \right)^2}{\omega_j^2}
  \label{eq:SEsignalPure}
\end{align}
for $\omega_j \gg  \gamma^{\prime}B_{\perp,j}^{(\U{spin})}$.
It is worth mentioning that since the signal does not  depend on the initial spin state $\ket{\pm}^{\rm{(spin)}}_j$, we obtain the same signal as Eq.(\ref{eq:SEsignalPure}) even when the initial spin state is completely mixed such as $\hat{\rho}_j^{\rm{(spin)}}=\frac{1}{2}\ket{+}\bra{+}_j^{\U{(spin)}}+\frac{1}{2}\ket{-}\bra{-}_j ^{\rm{(spin)}}$.
This shows that the polarization of the nuclear spins is not required to perform the NMR when we use the spin echo on the FQ.

We generalize this idea to the case of $M$ nuclear
spins.
The state before the readout step is described as
\begin{align}
  \ket{\psi_+ (\tau\u{DD})}=\frac{1}{\sqrt{2}}\ket{ 1}^{\U{(FQ)}} \otimes \prod_{j=1}^M \hat{U}_{a,j}\ket{+}_j^{\U{(spin)}} \nonumber \\
  +\frac{1}{\sqrt{2}}\ket{0}^{\U{(FQ)}} \otimes  \prod_{j=1}^M \hat{U}_{b,j} \ket{+}_j^{\U{(spin)}}.
\end{align}
By readout the state by $\hat{\mathcal{P}}_x$, we obtain
\begin{align}
  P\u{DD}&=\frac{1}{2} + \frac{1}{4}\prod_{j=1}^M {}_{\quad j}^{\U{(spin)}}\bra{+}(\hat{U}_{a,j}^{\dagger}\hat{U}_{b,j}
  +\hat{U}_{b,j}^{\dagger}\hat{U}_{a,j})\ket{+}_j^{\U{(spin)}}
  \nonumber \\
  &\simeq 1-\sum_{j=1}^{M}\left( \cos \frac{\omega_j \tau\u{DD}}{2}-1\right)^2 \frac{ \left(\gamma^{\prime} B_{\perp,j}^{(\U{spin})} \right)^2}{\omega_j^2},
  \label{eq:SinglePiSE}
\end{align}
for $\omega_j \gg \gamma^{\prime}B_{\perp,j}^{(\U{spin})}$.
Similar to the case of the FQ coupled with a single nuclear spin discussed above, we obtain the same signal even when the initial spin
state is completely mixed.
We can obtain the signal when we perform $\pi$ pulses $2n\!-\!1$ times, which corresponds to the case of the dynamical decoupling.
\begin{align}
  P\u{DD}&=\frac{1}{2} + \frac{1}{4} \nonumber \\
   & \times \prod_{j=1}^M {}_{\quad j}^{\U{(spin)}}\bra{+}\left(({\hat{U}_{a,j}^{\dagger}})^{n}(\hat{U}_{b,j})^{n}
  +(\hat{U}_{b,j}^{\dagger})^n (\hat{U}_{a,j})^n \right)\ket{+}_j^{\U{(spin)}}.
  \label{eq:SEsignalMixed}
\end{align}

In this section,
to understand the basic properties of the NMR with the FQ via the AC magnetic fields from the nuclear spins, we mainly discuss the simplest spin-echo case to perform a single $\pi$ pulse on the FQ,
while we show the detailed calculation of the case to perform the dynamical decoupling in Appendix~A.

As is the case with the Ramsey measurement with asymmetric driving, the nuclear spins are affected by low-frequency magnetic fields noise $\delta B^{(j)}\u{ex}$ from the environment.
To take into account this effect, we consider an ensemble average of the frequency with a Gaussian weight as follows.
\begin{align}
  \overline{P}\u{DD} &\simeq 1 - \sum_{j=1}^{M} \int_{-\infty}^{\infty}\frac{1}{\tilde{\Gamma}\sqrt{\pi}}\mathrm{e}^{-\frac{\delta\omega_j^2}{\tilde{\Gamma}^2}}\nonumber \\
  &\times \frac{ \left( \cos \frac{\omega_j \tau\u{DD}}{2}-1\right)^2 }{\omega_j^{2}} d(\delta\omega_j) (\gamma^{\prime}  B_{\perp,j}^{(\U{spin})})^2\nonumber \\
  &\simeq 1 - \left( \cos \frac{\omega \tau\u{DD}}{2}-1\right)^2 \sum_{j=1}^{M}\frac{(\gamma^{\prime}  B_{\perp,j}^{(\U{spin})})^2}{\omega^2},\nonumber \\
  \label{eq:low-frequencynoiseSE}
\end{align}
where we assume $\tilde{\Gamma}\ll \omega $.
So, the signal does not depend on the linewidth $\tilde{\Gamma }$ as long as the higher order terms of $|\tilde{\Gamma }/ \omega|$ is negligible.

We consider the dephasing of the FQ and an imperfect readout.
Due to the dephasing, the density matrix of the total system before the readout step is described as
\begin{align}
  \hat{\rho}\u{DD}&=\frac{1+\mathrm{e}^{-\Gamma\u{DD}^{\U{(FQ)}} \tau\u{DD}}}{2} \ket{\psi_{\pm} (\tau\u{DD})}\bra{\psi_{\pm} (\tau\u{DD})}\nonumber \\
  &+ \frac{1-\mathrm{e}^{-\Gamma\u{DD}^{\U{(FQ)}} \tau\u{DD}}}{2}( \hat{\sigma}_z^{(\U{FQ)}}\ket{\psi_{\pm} (\tau\u{DD})}\bra{\psi_{\pm} (\tau\u{DD})} \hat{\sigma}_z^{(\U{FQ)}}),
\end{align}
where $\Gamma\u{DD}^{\U{(FQ)}}=1/T_2$ is the dephasing rate of the FQ for dynamical decoupling.
Then, the signal with the imperfect readout is described as
\begin{align}
  &\overline{\tilde{P'}}\u{DD}\simeq \frac{1}{2} + V\mathrm{e}^{-\Gamma\u{DD}^{\U{(FQ)}}\tau\u{DD}}\nonumber \\
  &\times \left[ \frac{1}{2} -\left( \cos \frac{\omega \tau\u{DD}}{2}-1\right)^2 \sum_{j=1}^{M}
  \frac{(\gamma^{\prime} B_{\perp,j}^{(\U{spin})})^2}{\omega^2} \right].\nonumber \\
  \label{eq:PSEfinal}
\end{align}
Although the signal described here is the case of the spin echo, we show the signal form with the case of the general dynamical decoupling in the appendix A.

In our scheme, we measure an amplitude of AC magnetic fields generated by the nuclear spins.
According to the standard prescription of the quantum metrology~\cite{degan2017QuantumSensing}, we will consider the uncertainty of the estimation of the target fields as follows
\begin{align}
  \delta B_{\rm{AC}}
   =\frac{\sqrt{\overline{\tilde{P'}}\u{DD}\left(1-\overline{\tilde{P'}}\u{DD}\right)}}{\left|\frac{d\overline{\tilde{P'}}\u{DD}}{dB_{\rm{AC}} }\right|\sqrt{N}} \nonumber
   \label{eq:ImperfectionReadoutSE}
\end{align}
where $B_{\rm{AC}}=\sqrt{\sum_{j=1}^{M} \left( B_{\perp,j}^{(\U{spin})}\right)^2}$ denotes effective AC magnetic fields from the nuclear spins.
The interaction time $\tau\u{DD}$ is numerically determined to minimize this uncertainty $\delta B\u{AC}$.

\section{The detectable density and number of the nuclear  spins by NMR with the FQ}
\label{sec:CalculationSection}
To compare the performance of the two schemes (Ramsey measurement and dynamical decoupling), we will calculate the detectable density and the number of nuclear spins by using these two schemes.
To calculate the minimum detectable density of the nuclear spins, we consider a circumstance that a large spin sample containing nuclear spins is attached on the FQ with a minimum distance of $h$ as shown in Fig.~\ref{fig:FQsetup}.
On the other hand, to calculate the minimum detectable number of nuclear spins, we consider a spin sample whose size is smaller than the FQ.
For the calculations,
we set the temperature $T=20$~mK,
the qubit gap frequency $\Delta/2\pi=5.37$~GHz,
the frequency detuning $\epsilon/2\pi=0.112$~GHz,
the persistent current $I\u{FQ}=180$~nA,
the visibility $V=0.79$,
the repetition time $T\u{rep}\simeq 100~\mu$s,
the dephasing time for a Ramsey measurement $T^{\ast}_2=1~\mu$s,
the dephasing time for a dynamical decoupling with $n=1, 2, 4, 6, 8, 10$ is $T\u{2}=5.00, 6.63, 8.91, 10.8, 12.4, 13.6~\mu$s,
and the distance between the RF line and the FQ is set as $z\u{RF}=2~ \mu$m.
We use these parameters based on recent experimental results shown in~\cite{jonas2011noisespectroscopywithFQ}.
Also, we assume that the target nuclear spin is proton with a gyromagnetic ratio of $\gamma/2\pi\simeq42.6$~MHz/T, and the electric current for the RF driving strength is optimized.

\subsection{The minimum detectable density for NMR with the FQ}

\begin{figure}
\includegraphics[width=8cm]{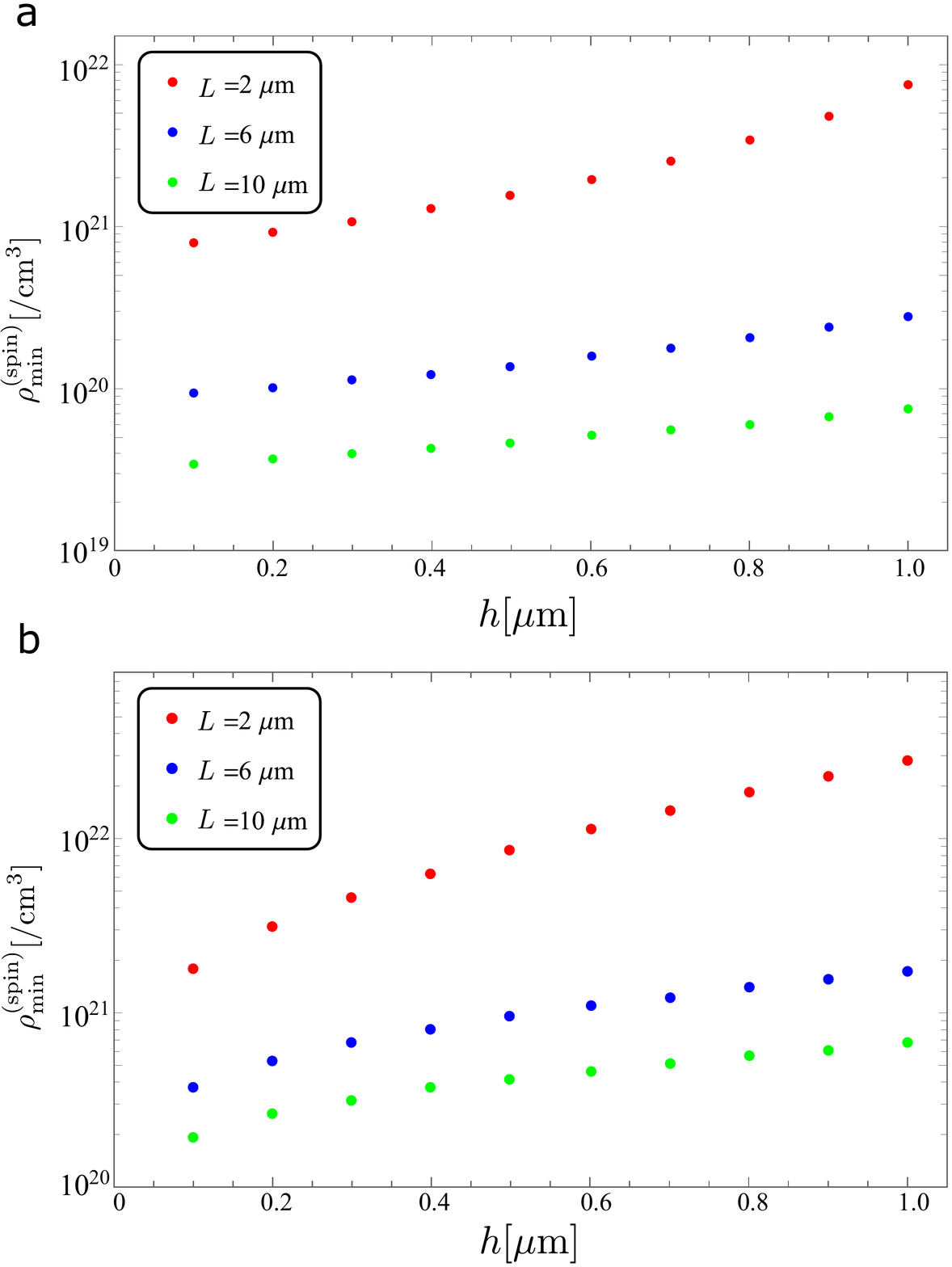}
\caption{The minimum detectable density against the distance $h$ between the spin sample and the FQ by using the Ramsey measurement with asymmetric driving~{\bf a} and
 the spin echo scheme~{\bf b}.
The red, blue, and green dots indicate the FQ
size $2~\mu$m, $6~\mu$m, and $10~\mu$m, respectively. }
\label{fig:TvsHight}
\end{figure}

\begin{figure}
\includegraphics[width=9cm, clip]{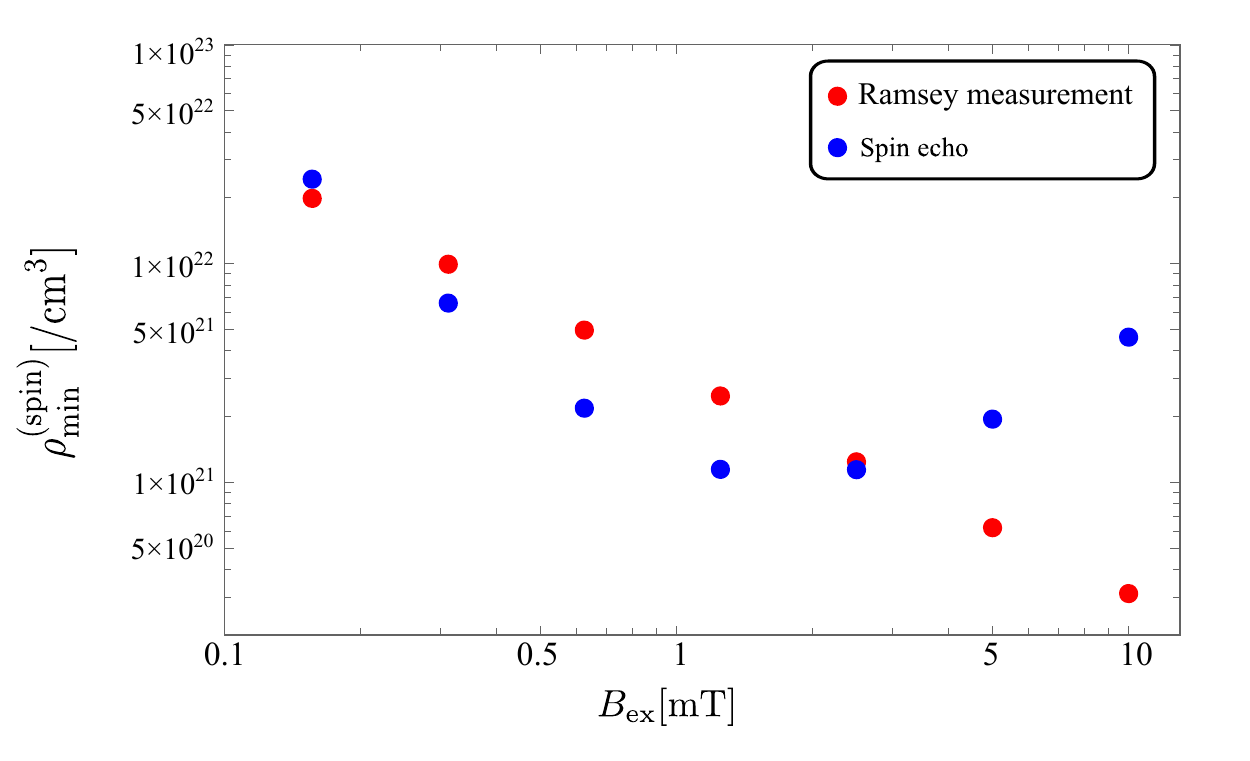}
\caption{
The minimum detectable density against the external magnetic field $B\u{ex}$ by using the Ramsey measurement with asymmetric driving and the spin echo scheme.
The red dots are the plot for the Ramsey measurement with asymmetric driving.
The blue dots are the plot for the spin echo scheme.
}
\label{fig:Mag}
\end{figure}

To calculate the minimum detectable density, we define signal-to-noise ratio~($SNR$).
In our NMR with the FQ, the signal is an amplitude of the effective magnetic field from the nuclear spins while the noise is the
uncertainty of the estimation.
When we fix the other parameters, both the signal and noise just depend on the density of the nuclear spins.
So we define the minimum detectable density $\rho_{\rm{min}}^{\rm{(spin)}}$  as to satisfy $B_{\rm{DC}}\left(\rho_{\rm{min}}^{\rm{(spin)}}\right)=\delta B_{\rm{DC}}\left(\rho_{\rm{min}}^{\rm{(spin)}}\right)$ for the Ramsey measurement with asymmetric driving and $B_{\rm{AC}}\left(\rho _{\rm{min}}^{\rm{(spin)}}\right)=\delta B_{\rm{AC}} \left(\rho _{\rm{min}}^{\rm{(spin)}}\right)$ for the spin echo scheme.

Also, we assume that the size of the spin sample containing the nuclear spins is $2.5L\times2.5L\times2L$.
Around the edge of the spin sample with this size, the Zeeman splitting of
the nuclear spin due to the magnetic fields from the FQ becomes 3 order
of magnitude smaller than the largest
Zeeman splitting of the nuclear spins located above the FQ line.
This means that, although a much larger spin sample such as a few millimeters is attached on the FQ in the real experiment~\cite{toida2017ESRwithFQ}, the size adopted in our calculation is large enough to consider the effective coupling between the FQ and the nuclear spins.

The numerical results for the minimum detectable density against the height $h$ and the size $L$ for these two schemes are shown in Fig.~\ref{fig:TvsHight}.
These results show that, with $h=0.1~\mu$m, the minimum detectable density with the Ramsey measurement with asymmetric driving is 2.28, 4.00, 5.56 times
smaller than that with the spin echo scheme for the size of 2~$\mu$m, 6~$\mu$m, 10~$\mu$m, respectively.
Also, these plots show that, to detect the smaller density spins, it is helpful to increase the size of the FQ and to close the distance between the FQ and the spin sample.
It should be noted that, in our calculation, we adopt a coherence time reported in the previous work~\cite{jonas2011noisespectroscopywithFQ} where the size of the FQ is around $2~\mu$m,
and we use the same coherence time of the FQ with different sizes for our calculations.
However, in real experiments, a larger FQ would show a shorter coherence time.
So, our calculation for the FQ with the size larger than $2~\mu$m would not be available in the current technology, but they show potentially achievable values in the near future technology that could provide us a larger FQ with the reasonably long coherence time.
It is worth mentioning that, in these calculations,  we set the external magnetic field as $B\u{ex}=4$~mT.
It is known that, if a magnetic field larger than a certain threshold strength is applied, the FQ could be damaged and would not work as a two-level system.
Such a threshold magnetic field strength strongly depends on the superconducting material, but it is typically around 4 mT for the FQ with four Josephson junctions~\cite{toida2017ESRwithFQ}.
So, in this paper, we mainly consider the applied magnetic fields around 4 mT.

Next, we calculate the minimum detectable density $\rho\u{min}^{\U{(spin)}}$ against the $B\u{ex}$ for the FQ of $L=2~\mu$m.
The numerical results are shown in Fig.~\ref{fig:Mag}.
These plots show that the $\rho^{\U{(spin)}}\u{min}$ with a Ramsey measurement with asymmetric driving is inversely proportional to the external magnetic field $B\u{ex}$.
This is because the signal of a Ramsey measurement with asymmetric driving is proportional to the polarization of the spins and the polarization linearly increases with the external magnetic fields in our parameter range.
The $\rho^{\U{(spin)}}\u{min}$ using spin echo scheme has the minimum
value at a certain value of $B\u{ex}$ for the following reasons.
When the magnetic field gets larger than that value, the signal
decreases due to a short interaction time between the FQ and nuclear spins. On the other hand, when the
magnetic field gets smaller than that value, the interaction time
becomes longer, however, the signal decreases due to the dephasing of the FQ [see Eq.(\ref{eq:PSEfinal})].
In this calculation, the $\rho^{\U{(spin)}}\u{min}$ using spin echo scheme takes the minimum value at the $B\u{ex} \simeq 1.8$~mT.
This behavior is quantitatively the same for different sizes of FQs.
The minimum detectable density with the spin echo scheme takes the minimum value of $\rho^{\rm{(spin)}}_{\rm{min}}\simeq 10^{21}/$cm$^3$ for $B\u{ex} \simeq 1.8$~mT where $\omega=4\pi /T_2$ is approximately satisfied.
This is consistent with the fact that the performance to sense the AC magnetic fields with a frequency of $\omega$ by using a qubit becomes optimized for $\omega \simeq 2\pi /\tau $ and $\tau \simeq T_2/2$~\cite{kitazawa2017vectormagnetsensing}.

\begin{figure}[t]
\includegraphics[width=9cm]{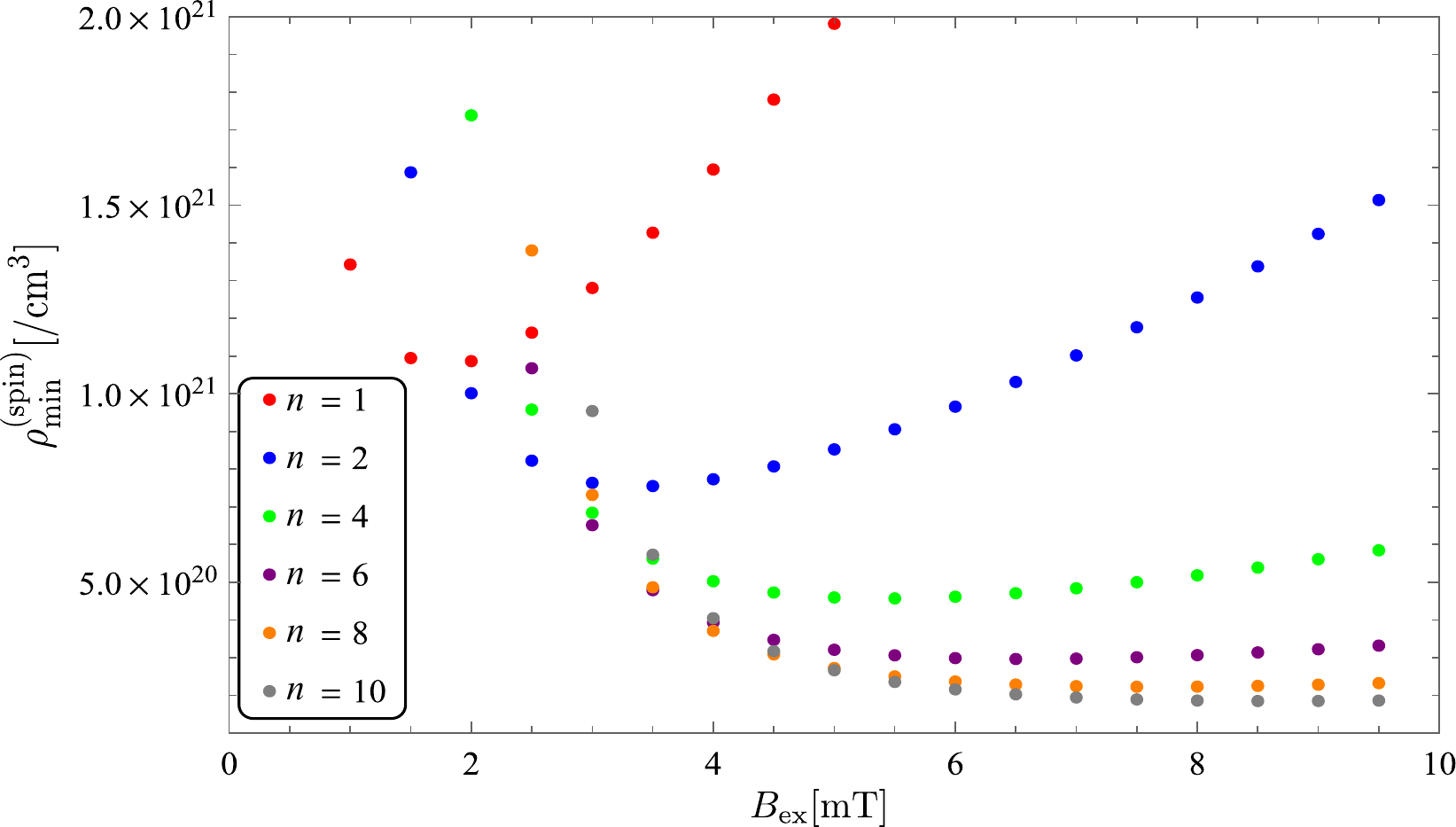}
\caption{The minimum detectable density against the external magnetic field $B\u{ex}$ and the echo times.
The red, blue, green, purple, orange and gray circle denotes the echo times $n=1, 2, 4, 6, 8, 10$.
}
\label{fig:DDvsBex}
\end{figure}

We also plot the magnetic field dependence of the $\rho\u{min}^{\U{(spin)}}$ for multiple $\pi$ pulses in Fig.~\ref{fig:DDvsBex}.
These calculations show that, by increasing both the number of the $\pi $ pulses and the strength of the applied magnetic fields, we can detect spins with a smaller density.
This comes from the fact that increasing the number of the $\pi$ pulses improves the coherence time, while the time interval between the $\pi$ pulses becomes shorter, which requires higher Larmor frequency of the nuclear spins
to synchronize with the $\pi$ pulse time interval on the FQ.
However, it is known that the FQ cannot stand the high external magnetic field $B\u{ex}$, as we discussed before.
Therefore, we consider a case of the applied magnetic field of $4$ mT that is close to the strongest applied magnetic fields with the FQ, and we find that the optimal number of the $\pi $ pulses with this magnetic fields
is $n=8$.

\subsection{The minimum detectable number of nuclear spins by NMR with the FQ}

\begin{figure}[t]
\includegraphics[width=8cm]{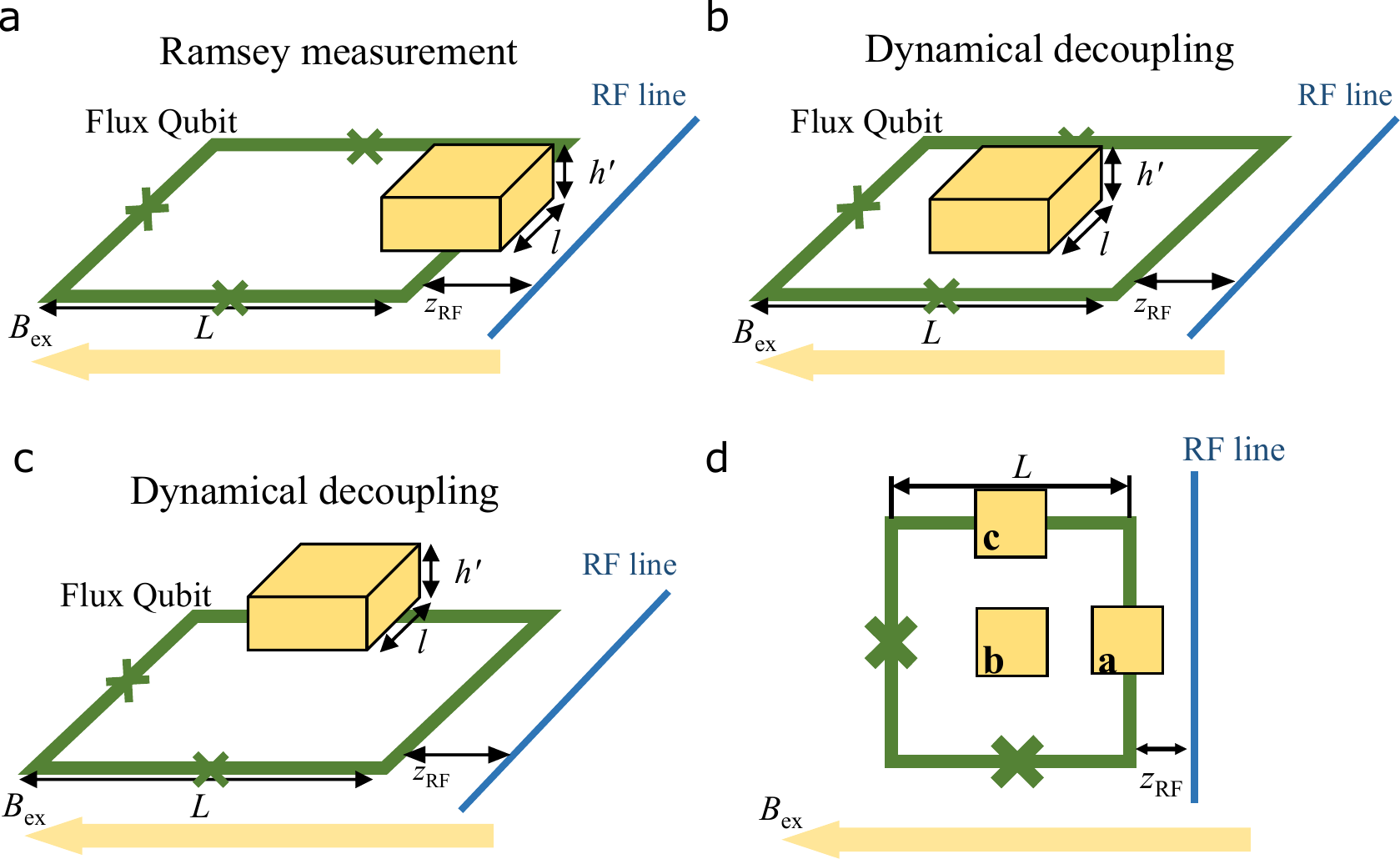}
\caption{The schematic for sensing the minimum detectable spin-number.
The size of the spin sample is $l\times l\times h'$.
The distance between the spin sample and the FQ is set as  $h=$0.1~$\mu$m.
The center of the spin sample is attached to the FQ in three ways.
{\bf a}, The spin sample center is put on the middle of the FQ line that is closest to the RF line.
{\bf b}, The spin sample center is set at the center of the FQ.
{\bf c},  The spin sample center is put on the middle of the FQ line that is orthogonal to the external magnetic field.
{\bf d}, The top view of the schematic setup {\bf a}, {\bf b} and {\bf c}.
}
\label{fig:Nspinsetup}
\end{figure}

We discuss how to estimate the minimum detectable number of nuclear spins $N\u{min}^{(\U{spin})}$.
In the current experiments, a large spin sample of millimeter size is attached on the FQ~\cite{toida2017ESRwithFQ}.
In this setup, the FQ has finite couplings with all nuclear spins in the large spin sample, thus, it is not straightforward to estimate the number of the detected spins.
If we naively sum up the number of the spins that have a finite coupling with the FQ, we need to consider every spin in the spin sample, which turned out to be quite large.
So, for the estimation of the $N\u{min}^{(\U{spin})}$, we will consider the case that the spin sample is as small as the FQ.
More specifically, we consider the setup as shown in Fig.~\ref{fig:Nspinsetup}.
Since the NMR signal comes from $B_{z,j}^{({\rm{spin}})}$ for a Ramsey measurement asymmetric driving while the NMR signal comes from $B_{\perp,j}^{({\rm{spin}})}$ for dynamical decoupling scheme, the optimized way to put the spin sample for each scheme should be different.
The size of the spin sample is $l\times l\times h'$ where $l$ ($h'$) denotes the width (height), and we set $h=0.1~\mu$m (the distance between the spin sample and the FQ) and $h'=0.1~\mu$m.
In this calculation, we assume that all nuclear spins are saturated with strong driving fields
for the Ramsey measurement.
For a given value of $l$, we calculate the minimum density
$\rho\u{min}^{\U{(spin)}}$ such that $SNR$ becomes unity in this
setup (similar to the case in the previous subsection), and the
 $N\u{min}^{(\U{spin})}$ can be calculated as $l\times l\times h'\times \rho\u{min}^{\U{(spin)}}$.

\begin{figure}[t]
\includegraphics[width=9cm]{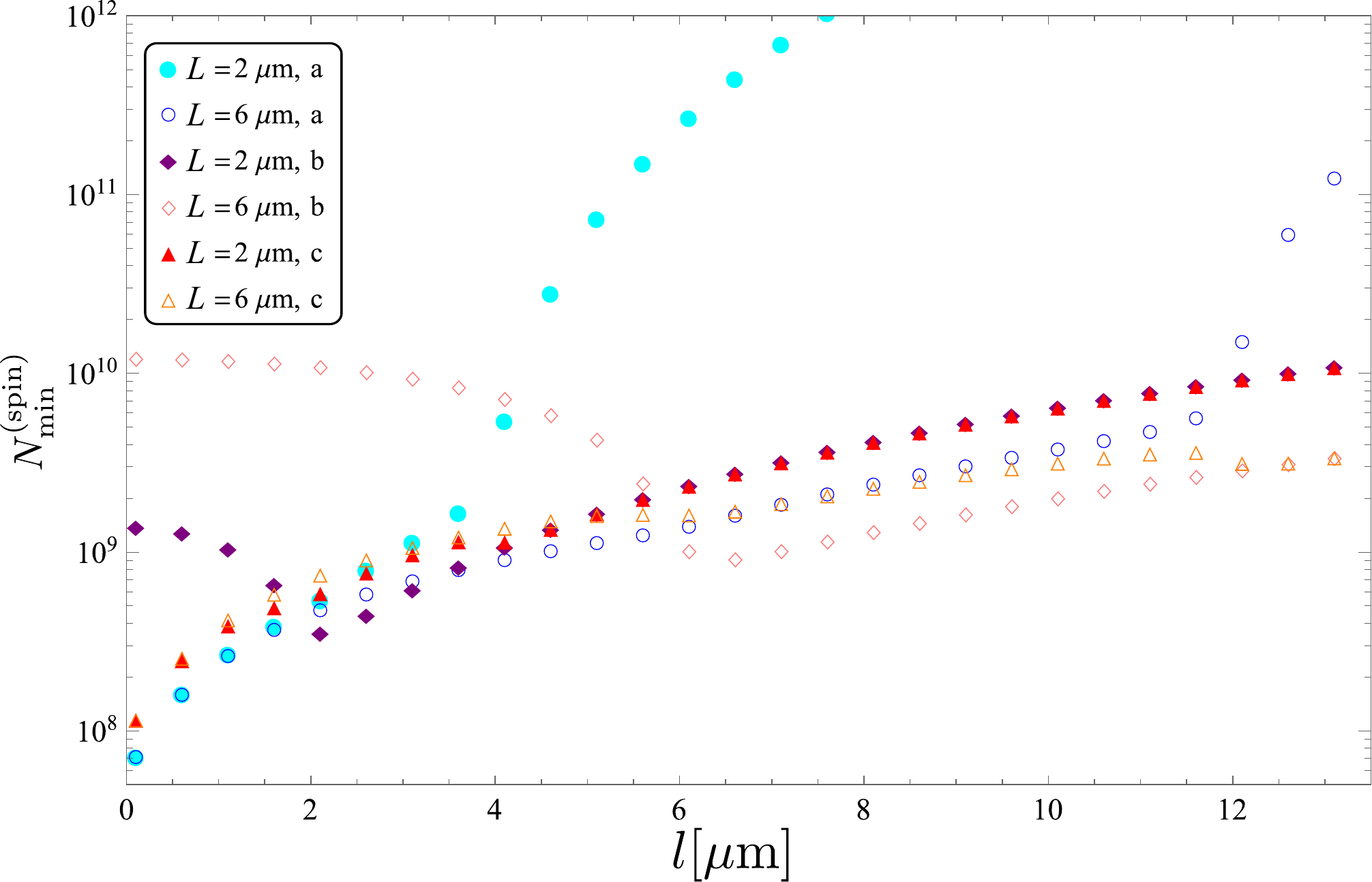}
\caption{The minimum detectable number of the nuclear spins against the size of the spin sample.
In the legends, {\bf a}, {\bf b}, and {\bf c} represent the sample configurations shown in Fig.~9.
The NMR with the setup {\bf a} and {\bf c} detects the minimum number of the nuclear spins when the size of the spin sample is much smaller than the size of the FQ,
while the NMR with the setup {\bf b} detects it when the size of the spin sample is comparable with that of the FQ.
}
\label{fig:NTTvsSPINECHO}
\end{figure}

The calculation results for the  $N\u{min}^{(\U{spin})}$ are shown in Fig.~\ref{fig:NTTvsSPINECHO}.
In this calculation, we set $B\u{ex}=4$~mT and use the dynamical decoupling with $n=8$.
When we use the FQ with the size of $L=2~\mu$m and the spin sample with the width $l$ of a few hundred nm, the $N\u{min}^{(\U{spin})}$ can be around $10^8$ either
by using the Ramsey measurement scheme in the setup {\bf a} or the dynamical decoupling scheme with $n=8$ in the setup {\bf b}.
The behaviors of the $N\u{min}^{\U{(spin)}}$ using a Ramsey measurement with asymmetric driving drastically change when the size of the spin sample is around $2L$.
Actually, $N\u{min}^{\U{(spin)}}$ for the spin sample with the size more than $2L$ increase more rapidly than that with the spin sample with the size less than $2L$.
This is because the spin sample with a size much larger than $2L$ contains many nuclear spins that are only weakly coupled with the FQ due to the long distance between them.
With the setup {\bf b}, the dynamical decoupling scheme can detect the smallest number of spins when the size of the spin sample is approximately equal to the size of the FQ.
This is reasonable because the spin sample with the size either much larger or smaller than $L$ makes the average FQ-spin coupling weaker for the dynamical decoupling scheme case.
The reason why the setup {\bf b} is better than the setup {\bf c} in the
dynamical decoupling scheme at $l\simeq L$  is that in the setup {\bf b},
the spin sample can exactly cover the FQ, which provides us with the optimized average coupling strength.
$N\u{min}^{\U{(spin)}}$ for other $B\u{ex}$ can be estimated by using both calculation results in Fig.~\ref{fig:Nspinsetup} and Fig.~\ref{fig:NTTvsSPINECHO}.

As we discussed above, we can approximately detect $10^8$ nuclear spins with our schemes in realistic conditions.
We compare this performance with that by using the other methods.
First, we compare the detectable number with the experimental results of the ESR with the FQ~\cite{toida2017ESRwithFQ}.
In this experiment, the FQ could detect the $\sim 400$ electron spins with an accumulation time of a second.
To take the ratio of the gyromagnetic ratio of electron and that of proton into consideration, it is presumed that the order of the detectable nuclear spins is around $10^8$, which is consistent with our numerical results.
An NMR using a conventional RF microcoil can detect $\sim 5\times10^{11}$ nuclear spins at room temperature and static magnetic field of 11.7~T with a 10 min acquisition time~\cite{subramanian1998RFmicrocoilNMR}.
The polarization of the nuclear spins is almost the same for this condition and our condition.
Compared to this number, the FQ can detect $10^3$ times smaller nuclear spins.

\section{conclusion}
In conclusion, we theoretically investigate the performance of the nuclear magnetic resonance~(NMR) when we use the superconducting flux qubit~(FQ) as the detector.
For NMR with the FQ, we discuss a Ramsey measurement
and a dynamical decoupling.
In the former scheme, we asymmetrically drive the nuclear spins by the RF signals, and the FQ detects the DC magnetic fields change due to the driving.
In the latter scheme, the FQ detects the AC magnetic field from the nuclear spins due to the Larmor precession.
We show that, in either case, the minimum detectable density (number) of the nuclear spins for the FQ is around $10^{21}/{\rm{cm}^3}$ ($10^8$) with an accumulation time of a second.
Our proposed NMR with the FQ is attractive because of the possibility to detect the nuclear spins at a local region ($\sim\mu$m) with low temperature ($\sim$ mK) and low magnetic fields ($\sim$ mT).

\section*{ACKNOWLEDGEMENTS}
This work was supported by CREST (JPMJCR1774 and JPMJCR1672), JST and Program for Leading Graduate Schools: Interactive Materials Science Cadet Program, and in part by MEXT Grants-in-Aid for Scientific Research on Innovative Areas ``Science of hybrid quantum systems'' (Grant No. 15H05870).

\section*{APPENDIX A: Signal of NMR using general dynamical decoupling scheme}
We derived the signal for NMR using dynamical decoupling with one $\pi$ pulse in Eq.~(\ref{eq:SinglePiSE}).
Here, we describe the signal for NMR using dynamical decoupling with $2n\!-\!1$ $\pi$ pulses.

When we use the dynamical decoupling with even $n$, the signal $P_{\U{DD}, n}$ in Eq.~(\ref{eq:SEsignalMixed}) becomes
\begin{align}
  P_{\U{DD}, n} &\simeq
  1 - \sum_{j=1}^{M} \left(2  \sum_{i=0}^{\frac{n}{2}-1} \cos \left( \frac{2i+1}{2}\tau\u{DD}\omega_j \right) \right)^2  \nonumber \\
   &\times \left( \cos \frac{\omega_j \tau\u{DD}}{2}-1\right)^2 \frac{ \left(\gamma^{\prime} B_{\perp,j}^{(\U{spin})} \right)^2}{\omega_j^2}
\end{align}
for $\omega_j \gg  \gamma^{\prime}B_{\perp,j}^{(\U{spin})}$.
When the $n$ is odd except for one, the signal $P_{\U{DD}, n}$ in Eq.~(\ref{eq:SEsignalMixed}) becomes
\begin{align}
  P_{\U{DD}, n} &\simeq
  1 - \sum_{j=1}^{M} \left(1 + 2  \sum_{i=0}^{\frac{n-1}{2}-1} \cos \left( (i+1)\tau\u{DD}\omega_j \right) \right)^2 \nonumber \\
  &\times \left( \cos \frac{\omega_j \tau\u{DD}}{2}-1\right)^2 \frac{ \left(\gamma^{\prime} B_{\perp,j}^{(\U{spin})} \right)^2}{\omega_j^2}
\end{align}
for $\omega_j \gg  \gamma^{\prime}B_{\perp,j}^{(\U{spin})}$.
Similar to the case of the FQ coupled with a single nuclear spin discussed above, we obtain the same signal even when the initial spin
state is completely mixed.

The signal considering the effect of low-frequency magnetic field noise, dephasing of the FQ and the imperfect readout, which is Eq.~(\ref{eq:PSEfinal}) for the case of $n=1$, is
\begin{align}
  \overline{\tilde{P'}}_{\U{DD},n}\simeq & \frac{1}{2} + V\mathrm{e}^{-\Gamma\u{DD}^{\U{(FQ)}}\tau\u{DD}}\nonumber \\
  &\times \Biggl[ \frac{1}{2} - \left(2  \sum_{i=0}^{\frac{n}{2}-1} \cos \left( \frac{2i+1}{2}\tau\u{DD}\omega \right) \right)^2 \nonumber \\
  & \times \left( \cos \frac{\omega \tau\u{DD}}{2}-1\right)^2 \sum_{j=1}^{M}
  \frac{\left(\gamma^{\prime} B_{\perp,j}^{(\U{spin})}\right)^2}{\omega^2} \Biggr],
\end{align}
for $n$ is even.
Regarding the dynamical decoupling with odd $n$, the signal considering those effect is
\begin{align}
  \overline{\tilde{P'}}_{\U{DD},n}\simeq & \frac{1}{2} + V\mathrm{e}^{-\Gamma\u{DD}^{\U{(FQ)}}\tau\u{DD}}\nonumber \\
  &\times \Biggl[ \frac{1}{2} - \left(1 + 2  \sum_{i=0}^{\frac{n-1}{2}-1} \cos \left( (i+1)\tau\u{DD}\omega \right) \right)^2 \nonumber \\
  & \times \left( \cos \frac{\omega \tau\u{DD}}{2}-1\right)^2 \sum_{j=1}^{M}
  \frac{\left(\gamma^{\prime} B_{\perp,j}^{(\U{spin})}\right)^2}{\omega^2} \Biggr].
\end{align}


\begin{thebibliography}{55}
\expandafter\ifx\csname natexlab\endcsname\relax\def\natexlab#1{#1}\fi
\expandafter\ifx\csname bibnamefont\endcsname\relax
  \def\bibnamefont#1{#1}\fi
\expandafter\ifx\csname bibfnamefont\endcsname\relax
  \def\bibfnamefont#1{#1}\fi
\expandafter\ifx\csname citenamefont\endcsname\relax
  \def\citenamefont#1{#1}\fi
\expandafter\ifx\csname url\endcsname\relax
  \def\url#1{\texttt{#1}}\fi
\expandafter\ifx\csname urlprefix\endcsname\relax\def\urlprefix{URL }\fi
\providecommand{\bibinfo}[2]{#2}
\providecommand{\eprint}[2][]{\url{#2}}

\bibitem[{\citenamefont{R.~Ernst et~al.}(1988)\citenamefont{R.~Ernst,
  Bodenhausen, and Wokaun}}]{ernst1988principlesofnmr}
\bibinfo{author}{\bibfnamefont{R.}~\bibnamefont{R.~Ernst}},
  \bibinfo{author}{\bibfnamefont{G.}~\bibnamefont{Bodenhausen}},
  \bibnamefont{and} \bibinfo{author}{\bibfnamefont{A.}~\bibnamefont{Wokaun}},
  \emph{\bibinfo{title}{Principles of nuclear magnetic resonance in one and two
  dimensions}} (\bibinfo{publisher}{Oxford University Press},
  \bibinfo{year}{1988}).

\bibitem[{\citenamefont{Cavalli et~al.}(2007)\citenamefont{Cavalli, Salvatella,
  Dobson, and Vendruscolo}}]{cavalli2007proteinstructuredetermination}
\bibinfo{author}{\bibfnamefont{A.}~\bibnamefont{Cavalli}},
  \bibinfo{author}{\bibfnamefont{X.}~\bibnamefont{Salvatella}},
  \bibinfo{author}{\bibfnamefont{C.~M.} \bibnamefont{Dobson}},
  \bibnamefont{and}
  \bibinfo{author}{\bibfnamefont{M.}~\bibnamefont{Vendruscolo}},
  \bibinfo{journal}{Proceedings of the National Academy of Sciences}
  \textbf{\bibinfo{volume}{104}}, \bibinfo{pages}{9615} (\bibinfo{year}{2007}).

\bibitem[{\citenamefont{Aguayo et~al.}(1986)\citenamefont{Aguayo, Blackband,
  Schoeniger, Mattingly, and Hintermann}}]{abuayo1986NMRimaging}
\bibinfo{author}{\bibfnamefont{J.~B.} \bibnamefont{Aguayo}},
  \bibinfo{author}{\bibfnamefont{S.~J.} \bibnamefont{Blackband}},
  \bibinfo{author}{\bibfnamefont{J.}~\bibnamefont{Schoeniger}},
  \bibinfo{author}{\bibfnamefont{M.~A.} \bibnamefont{Mattingly}},
  \bibnamefont{and}
  \bibinfo{author}{\bibfnamefont{M.}~\bibnamefont{Hintermann}},
  \bibinfo{journal}{Nature} \textbf{\bibinfo{volume}{322}},
  \bibinfo{pages}{190} (\bibinfo{year}{1986}).

\bibitem[{\citenamefont{W\"{u}thrich}(1986)}]{wuthrich1986NMRofPandNA}
\bibinfo{author}{\bibfnamefont{K.}~\bibnamefont{W\"{u}thrich}},
  \emph{\bibinfo{title}{NMR of proteins and nucleic acids}}
  (\bibinfo{publisher}{John Wiley, Sons New York}, \bibinfo{year}{1986}).

\bibitem[{\citenamefont{Overhauser}(1953)}]{Overhauser53}
\bibinfo{author}{\bibfnamefont{A.~W.} \bibnamefont{Overhauser}},
  \bibinfo{journal}{Phys. Rev.} \textbf{\bibinfo{volume}{92}},
  \bibinfo{pages}{411} (\bibinfo{year}{1953}).

\bibitem[{\citenamefont{Augustine et~al.}(1998)\citenamefont{Augustine,
  TonThat, and Clarke}}]{augustine1998SQUIDNMR}
\bibinfo{author}{\bibfnamefont{M.~P.} \bibnamefont{Augustine}},
  \bibinfo{author}{\bibfnamefont{D.~M.} \bibnamefont{TonThat}},
  \bibnamefont{and} \bibinfo{author}{\bibfnamefont{J.}~\bibnamefont{Clarke}},
  \bibinfo{journal}{Solid State Nuclear Magnetic Resonance}
  \textbf{\bibinfo{volume}{11}}, \bibinfo{pages}{139 } (\bibinfo{year}{1998}),
  ISSN \bibinfo{issn}{0926-2040}.

\bibitem[{\citenamefont{Poggio and Degen}(2010)}]{0957-4484-21-34-342001}
\bibinfo{author}{\bibfnamefont{M.}~\bibnamefont{Poggio}} \bibnamefont{and}
  \bibinfo{author}{\bibfnamefont{C.~L.} \bibnamefont{Degen}},
  \bibinfo{journal}{Nanotechnology} \textbf{\bibinfo{volume}{21}},
  \bibinfo{pages}{342001} (\bibinfo{year}{2010}).

\bibitem[{\citenamefont{Maguire et~al.}(2007)\citenamefont{Maguire, Chuang,
  Zhang, and Gershenfeld}}]{Maguire2007}
\bibinfo{author}{\bibfnamefont{Y.}~\bibnamefont{Maguire}},
  \bibinfo{author}{\bibfnamefont{I.~L.} \bibnamefont{Chuang}},
  \bibinfo{author}{\bibfnamefont{S.}~\bibnamefont{Zhang}}, \bibnamefont{and}
  \bibinfo{author}{\bibfnamefont{N.}~\bibnamefont{Gershenfeld}},
  \bibinfo{journal}{Proceedings of the National Academy of Sciences}
  \textbf{\bibinfo{volume}{104}}, \bibinfo{pages}{9198} (\bibinfo{year}{2007}),
  ISSN \bibinfo{issn}{0027-8424}.

\bibitem[{\citenamefont{Suefke et~al.}(2015)\citenamefont{Suefke, Liebisch,
  Bl\"{u}mich, and Appelt}}]{martin2015EHQFresonator}
\bibinfo{author}{\bibfnamefont{M.}~\bibnamefont{Suefke}},
  \bibinfo{author}{\bibfnamefont{A.}~\bibnamefont{Liebisch}},
  \bibinfo{author}{\bibfnamefont{B.}~\bibnamefont{Bl\"{u}mich}},
  \bibnamefont{and} \bibinfo{author}{\bibfnamefont{S.}~\bibnamefont{Appelt}},
  \bibinfo{journal}{Nature Physics} \textbf{\bibinfo{volume}{11}},
  \bibinfo{pages}{767} (\bibinfo{year}{2015}).

\bibitem[{\citenamefont{Kolkowitz et~al.}(2012)\citenamefont{Kolkowitz,
  Unterreithmeier, Bennett, and
  Lukin}}]{kolkowitz2012nmrsensingSingleElectronSpin}
\bibinfo{author}{\bibfnamefont{S.}~\bibnamefont{Kolkowitz}},
  \bibinfo{author}{\bibfnamefont{Q.~P.} \bibnamefont{Unterreithmeier}},
  \bibinfo{author}{\bibfnamefont{S.~D.} \bibnamefont{Bennett}},
  \bibnamefont{and} \bibinfo{author}{\bibfnamefont{M.~D.} \bibnamefont{Lukin}},
  \bibinfo{journal}{Phys. Rev. Lett.} \textbf{\bibinfo{volume}{109}},
  \bibinfo{pages}{137601} (\bibinfo{year}{2012}).

\bibitem[{\citenamefont{Mamin et~al.}(2013)\citenamefont{Mamin, Kim, Sherwood,
  Rettner, Ohno, Awschalom, and Rugar}}]{mamin2013NanoscaleNMR}
\bibinfo{author}{\bibfnamefont{H.~J.} \bibnamefont{Mamin}},
  \bibinfo{author}{\bibfnamefont{M.}~\bibnamefont{Kim}},
  \bibinfo{author}{\bibfnamefont{M.~H.} \bibnamefont{Sherwood}},
  \bibinfo{author}{\bibfnamefont{C.~T.} \bibnamefont{Rettner}},
  \bibinfo{author}{\bibfnamefont{K.}~\bibnamefont{Ohno}},
  \bibinfo{author}{\bibfnamefont{D.~D.} \bibnamefont{Awschalom}},
  \bibnamefont{and} \bibinfo{author}{\bibfnamefont{D.}~\bibnamefont{Rugar}},
  \bibinfo{journal}{Science} \textbf{\bibinfo{volume}{339}},
  \bibinfo{pages}{557} (\bibinfo{year}{2013}).

\bibitem[{\citenamefont{Staudacher et~al.}(2013)\citenamefont{Staudacher, Shi,
  Pezzagna, Meijer, Du, Meriles, Reinhard, and
  Wrachtrup}}]{staudacher2013NMRon5nanometer}
\bibinfo{author}{\bibfnamefont{T.}~\bibnamefont{Staudacher}},
  \bibinfo{author}{\bibfnamefont{F.}~\bibnamefont{Shi}},
  \bibinfo{author}{\bibfnamefont{S.}~\bibnamefont{Pezzagna}},
  \bibinfo{author}{\bibfnamefont{J.}~\bibnamefont{Meijer}},
  \bibinfo{author}{\bibfnamefont{J.}~\bibnamefont{Du}},
  \bibinfo{author}{\bibfnamefont{C.~A.} \bibnamefont{Meriles}},
  \bibinfo{author}{\bibfnamefont{F.}~\bibnamefont{Reinhard}}, \bibnamefont{and}
  \bibinfo{author}{\bibfnamefont{J.}~\bibnamefont{Wrachtrup}},
  \bibinfo{journal}{Science} \textbf{\bibinfo{volume}{339}},
  \bibinfo{pages}{561} (\bibinfo{year}{2013}).

\bibitem[{\citenamefont{Balasubramanian
  et~al.}(2009)\citenamefont{Balasubramanian, Neumann, Twitchen, Markham,
  Kolesov, Mizuochi, Isoya, Achard, Beck, Tissler
  et~al.}}]{balasubramanian2009ultralong}
\bibinfo{author}{\bibfnamefont{G.}~\bibnamefont{Balasubramanian}},
  \bibinfo{author}{\bibfnamefont{P.}~\bibnamefont{Neumann}},
  \bibinfo{author}{\bibfnamefont{D.}~\bibnamefont{Twitchen}},
  \bibinfo{author}{\bibfnamefont{M.}~\bibnamefont{Markham}},
  \bibinfo{author}{\bibfnamefont{R.}~\bibnamefont{Kolesov}},
  \bibinfo{author}{\bibfnamefont{N.}~\bibnamefont{Mizuochi}},
  \bibinfo{author}{\bibfnamefont{J.}~\bibnamefont{Isoya}},
  \bibinfo{author}{\bibfnamefont{J.}~\bibnamefont{Achard}},
  \bibinfo{author}{\bibfnamefont{J.}~\bibnamefont{Beck}},
  \bibinfo{author}{\bibfnamefont{J.}~\bibnamefont{Tissler}},
  \bibnamefont{et~al.}, \bibinfo{journal}{Nature materials}
  \textbf{\bibinfo{volume}{8}}, \bibinfo{pages}{383} (\bibinfo{year}{2009}).

\bibitem[{\citenamefont{Mizuochi et~al.}(2009)\citenamefont{Mizuochi, Neumann,
  Rempp, Beck, Jacques, Siyushev, Nakamura, Twitchen, Watanabe, Yamasaki
  et~al.}}]{mizuochi2009coherence}
\bibinfo{author}{\bibfnamefont{N.}~\bibnamefont{Mizuochi}},
  \bibinfo{author}{\bibfnamefont{P.}~\bibnamefont{Neumann}},
  \bibinfo{author}{\bibfnamefont{F.}~\bibnamefont{Rempp}},
  \bibinfo{author}{\bibfnamefont{J.}~\bibnamefont{Beck}},
  \bibinfo{author}{\bibfnamefont{V.}~\bibnamefont{Jacques}},
  \bibinfo{author}{\bibfnamefont{P.}~\bibnamefont{Siyushev}},
  \bibinfo{author}{\bibfnamefont{K.}~\bibnamefont{Nakamura}},
  \bibinfo{author}{\bibfnamefont{D.}~\bibnamefont{Twitchen}},
  \bibinfo{author}{\bibfnamefont{H.}~\bibnamefont{Watanabe}},
  \bibinfo{author}{\bibfnamefont{S.}~\bibnamefont{Yamasaki}},
  \bibnamefont{et~al.}, \bibinfo{journal}{Phys. Rev. B}
  \textbf{\bibinfo{volume}{80}}, \bibinfo{pages}{041201}
  (\bibinfo{year}{2009}).

\bibitem[{\citenamefont{Bar-Gill et~al.}(2013)\citenamefont{Bar-Gill, Pham,
  Jarmola, Budker, and Walsworth}}]{bar-gill2013onesecondcoherence}
\bibinfo{author}{\bibfnamefont{N.}~\bibnamefont{Bar-Gill}},
  \bibinfo{author}{\bibfnamefont{L.}~\bibnamefont{Pham}},
  \bibinfo{author}{\bibfnamefont{A.}~\bibnamefont{Jarmola}},
  \bibinfo{author}{\bibfnamefont{D.}~\bibnamefont{Budker}}, \bibnamefont{and}
  \bibinfo{author}{\bibfnamefont{R.}~\bibnamefont{Walsworth}},
  \bibinfo{journal}{Nature Communications} \textbf{\bibinfo{volume}{4}},
  \bibinfo{pages}{1743} (\bibinfo{year}{2013}).

\bibitem[{\citenamefont{M\"{u}ller et~al.}(2014)\citenamefont{M\"{u}ller, Kong,
  Cai, Melentijevi\'c, Stacey, Markham, Twitchen, Isoya, Pezzagna, Meijer
  et~al.}}]{muller2014NMRwithsinglespin}
\bibinfo{author}{\bibfnamefont{C.}~\bibnamefont{M\"{u}ller}},
  \bibinfo{author}{\bibfnamefont{X.}~\bibnamefont{Kong}},
  \bibinfo{author}{\bibfnamefont{J.-M.} \bibnamefont{Cai}},
  \bibinfo{author}{\bibfnamefont{K.}~\bibnamefont{Melentijevi\'c}},
  \bibinfo{author}{\bibfnamefont{A.}~\bibnamefont{Stacey}},
  \bibinfo{author}{\bibfnamefont{M.}~\bibnamefont{Markham}},
  \bibinfo{author}{\bibfnamefont{D.}~\bibnamefont{Twitchen}},
  \bibinfo{author}{\bibfnamefont{J.}~\bibnamefont{Isoya}},
  \bibinfo{author}{\bibfnamefont{S.}~\bibnamefont{Pezzagna}},
  \bibinfo{author}{\bibfnamefont{J.}~\bibnamefont{Meijer}},
  \bibnamefont{et~al.}, \bibinfo{journal}{Nature communications}
  \textbf{\bibinfo{volume}{5}}, \bibinfo{pages}{4703} (\bibinfo{year}{2014}).

\bibitem[{\citenamefont{Degen et~al.}(2017)\citenamefont{Degen, Reinhard, and
  Cappellaro}}]{degan2017QuantumSensing}
\bibinfo{author}{\bibfnamefont{C.~L.} \bibnamefont{Degen}},
  \bibinfo{author}{\bibfnamefont{F.}~\bibnamefont{Reinhard}}, \bibnamefont{and}
  \bibinfo{author}{\bibfnamefont{P.}~\bibnamefont{Cappellaro}},
  \bibinfo{journal}{Reviews Of Modern Physics} \textbf{\bibinfo{volume}{89}},
  \bibinfo{pages}{035002} (\bibinfo{year}{2017}).

\bibitem[{\citenamefont{Dolde et~al.}(2013)\citenamefont{Dolde, Jakobi,
  Naydenov, Zhao, Pezzagna, Trautmann, AMeijer, Neumann, Jelezko, and
  Wrachtrup}}]{Dolde2013EntanglementNVcenterExperiment}
\bibinfo{author}{\bibfnamefont{F.}~\bibnamefont{Dolde}},
  \bibinfo{author}{\bibfnamefont{I.}~\bibnamefont{Jakobi}},
  \bibinfo{author}{\bibfnamefont{B.}~\bibnamefont{Naydenov}},
  \bibinfo{author}{\bibfnamefont{N.}~\bibnamefont{Zhao}},
  \bibinfo{author}{\bibfnamefont{S.}~\bibnamefont{Pezzagna}},
  \bibinfo{author}{\bibfnamefont{C.}~\bibnamefont{Trautmann}},
  \bibinfo{author}{\bibfnamefont{J.}~\bibnamefont{AMeijer}},
  \bibinfo{author}{\bibfnamefont{P.}~\bibnamefont{Neumann}},
  \bibinfo{author}{\bibfnamefont{F.}~\bibnamefont{Jelezko}}, \bibnamefont{and}
  \bibinfo{author}{\bibfnamefont{J.}~\bibnamefont{Wrachtrup}},
  \bibinfo{journal}{Nature Physics} \textbf{\bibinfo{volume}{9}}
  (\bibinfo{year}{2013}).

\bibitem[{\citenamefont{Yao et~al.}(2012)\citenamefont{Yao, Jiang, Gorshkov,
  Maurer, Giedke, Cirac, and Lukin}}]{Yao2012EntanglementNVcenterTheory}
\bibinfo{author}{\bibfnamefont{N.}~\bibnamefont{Yao}},
  \bibinfo{author}{\bibfnamefont{L.}~\bibnamefont{Jiang}},
  \bibinfo{author}{\bibfnamefont{A.}~\bibnamefont{Gorshkov}},
  \bibinfo{author}{\bibfnamefont{P.}~\bibnamefont{Maurer}},
  \bibinfo{author}{\bibfnamefont{G.}~\bibnamefont{Giedke}},
  \bibinfo{author}{\bibfnamefont{J.}~\bibnamefont{Cirac}}, \bibnamefont{and}
  \bibinfo{author}{\bibfnamefont{M.}~\bibnamefont{Lukin}},
  \bibinfo{journal}{Nature Communications} \textbf{\bibinfo{volume}{3}}
  (\bibinfo{year}{2012}).

\bibitem[{\citenamefont{Orlando et~al.}(1999)\citenamefont{Orlando, Mooij,
  Tian, van~der Wal, Levitov, Lloyd, and Mazo}}]{orlando1999firstFQtheory}
\bibinfo{author}{\bibfnamefont{T.~P.} \bibnamefont{Orlando}},
  \bibinfo{author}{\bibfnamefont{J.~E.} \bibnamefont{Mooij}},
  \bibinfo{author}{\bibfnamefont{L.}~\bibnamefont{Tian}},
  \bibinfo{author}{\bibfnamefont{C.~H.} \bibnamefont{van~der Wal}},
  \bibinfo{author}{\bibfnamefont{L.~S.} \bibnamefont{Levitov}},
  \bibinfo{author}{\bibfnamefont{S.}~\bibnamefont{Lloyd}}, \bibnamefont{and}
  \bibinfo{author}{\bibfnamefont{J.~J.} \bibnamefont{Mazo}},
  \bibinfo{journal}{Phys. Rev. B} \textbf{\bibinfo{volume}{60}},
  \bibinfo{pages}{15398} (\bibinfo{year}{1999}).

\bibitem[{\citenamefont{Chiorescu et~al.}(2003)\citenamefont{Chiorescu,
  Nakamura, Harmans, and Mooij}}]{Chiorescu2003CoherentOscillations}
\bibinfo{author}{\bibfnamefont{I.}~\bibnamefont{Chiorescu}},
  \bibinfo{author}{\bibfnamefont{Y.}~\bibnamefont{Nakamura}},
  \bibinfo{author}{\bibfnamefont{C.~J. P.~M.} \bibnamefont{Harmans}},
  \bibnamefont{and} \bibinfo{author}{\bibfnamefont{J.~E.} \bibnamefont{Mooij}},
  \bibinfo{journal}{Science} \textbf{\bibinfo{volume}{299}},
  \bibinfo{pages}{1869} (\bibinfo{year}{2003}), ISSN \bibinfo{issn}{0036-8075}.

\bibitem[{\citenamefont{Stern et~al.}(2014)\citenamefont{Stern, Catelani, Kubo,
  Grezes, Bienfait, Vion, Esteve, and Bertet}}]{stern2014flux3Dcavity}
\bibinfo{author}{\bibfnamefont{M.}~\bibnamefont{Stern}},
  \bibinfo{author}{\bibfnamefont{G.}~\bibnamefont{Catelani}},
  \bibinfo{author}{\bibfnamefont{Y.}~\bibnamefont{Kubo}},
  \bibinfo{author}{\bibfnamefont{C.}~\bibnamefont{Grezes}},
  \bibinfo{author}{\bibfnamefont{A.}~\bibnamefont{Bienfait}},
  \bibinfo{author}{\bibfnamefont{D.}~\bibnamefont{Vion}},
  \bibinfo{author}{\bibfnamefont{D.}~\bibnamefont{Esteve}}, \bibnamefont{and}
  \bibinfo{author}{\bibfnamefont{P.}~\bibnamefont{Bertet}},
  \bibinfo{journal}{Phys. Rev. Lett.} \textbf{\bibinfo{volume}{113}},
  \bibinfo{pages}{123601} (\bibinfo{year}{2014}).

\bibitem[{\citenamefont{Yan et~al.}(2016)\citenamefont{Yan, Gustavsson, Kama,
  Birenbaum, Sears, Hover, Gudmundsen, Rosenberg, Samach, Weber
  et~al.}}]{yan2016FQrevisited}
\bibinfo{author}{\bibfnamefont{F.}~\bibnamefont{Yan}},
  \bibinfo{author}{\bibfnamefont{S.}~\bibnamefont{Gustavsson}},
  \bibinfo{author}{\bibfnamefont{A.}~\bibnamefont{Kama}},
  \bibinfo{author}{\bibfnamefont{J.}~\bibnamefont{Birenbaum}},
  \bibinfo{author}{\bibfnamefont{A.~P.} \bibnamefont{Sears}},
  \bibinfo{author}{\bibfnamefont{D.}~\bibnamefont{Hover}},
  \bibinfo{author}{\bibfnamefont{T.~J.} \bibnamefont{Gudmundsen}},
  \bibinfo{author}{\bibfnamefont{D.}~\bibnamefont{Rosenberg}},
  \bibinfo{author}{\bibfnamefont{G.}~\bibnamefont{Samach}},
  \bibinfo{author}{\bibfnamefont{S.}~\bibnamefont{Weber}},
  \bibnamefont{et~al.}, \bibinfo{journal}{Nature communications}
  \textbf{\bibinfo{volume}{7}}, \bibinfo{pages}{12964} (\bibinfo{year}{2016}).

\bibitem[{\citenamefont{Chiorescu et~al.}(2004)\citenamefont{Chiorescu, Bertet,
  Semba, Nakamura, Harmans, and Mooij}}]{chiorescu2004coherent}
\bibinfo{author}{\bibfnamefont{I.}~\bibnamefont{Chiorescu}},
  \bibinfo{author}{\bibfnamefont{P.}~\bibnamefont{Bertet}},
  \bibinfo{author}{\bibfnamefont{K.}~\bibnamefont{Semba}},
  \bibinfo{author}{\bibfnamefont{Y.}~\bibnamefont{Nakamura}},
  \bibinfo{author}{\bibfnamefont{C.}~\bibnamefont{Harmans}}, \bibnamefont{and}
  \bibinfo{author}{\bibfnamefont{J.}~\bibnamefont{Mooij}},
  \bibinfo{journal}{Nature} \textbf{\bibinfo{volume}{431}},
  \bibinfo{pages}{159} (\bibinfo{year}{2004}).

\bibitem[{\citenamefont{Plantenberg et~al.}(2007)\citenamefont{Plantenberg,
  De~Groot, Harmans, and Mooij}}]{plantenberg2007demonstration}
\bibinfo{author}{\bibfnamefont{J.}~\bibnamefont{Plantenberg}},
  \bibinfo{author}{\bibfnamefont{P.}~\bibnamefont{De~Groot}},
  \bibinfo{author}{\bibfnamefont{C.}~\bibnamefont{Harmans}}, \bibnamefont{and}
  \bibinfo{author}{\bibfnamefont{J.}~\bibnamefont{Mooij}},
  \bibinfo{journal}{Nature} \textbf{\bibinfo{volume}{447}},
  \bibinfo{pages}{836} (\bibinfo{year}{2007}).

\bibitem[{\citenamefont{Lupa{\c{s}}cu et~al.}(2004)\citenamefont{Lupa{\c{s}}cu,
  Verwijs, Schouten, Harmans, and Mooij}}]{lupacscu2004nondestructive}
\bibinfo{author}{\bibfnamefont{A.}~\bibnamefont{Lupa{\c{s}}cu}},
  \bibinfo{author}{\bibfnamefont{C.}~\bibnamefont{Verwijs}},
  \bibinfo{author}{\bibfnamefont{R.}~\bibnamefont{Schouten}},
  \bibinfo{author}{\bibfnamefont{C.}~\bibnamefont{Harmans}}, \bibnamefont{and}
  \bibinfo{author}{\bibfnamefont{J.}~\bibnamefont{Mooij}},
  \bibinfo{journal}{Phys. Rev. Lett.} \textbf{\bibinfo{volume}{93}},
  \bibinfo{pages}{177006} (\bibinfo{year}{2004}).

\bibitem[{\citenamefont{Lupa{\c{s}}cu et~al.}(2006)\citenamefont{Lupa{\c{s}}cu,
  Driessen, Roschier, Harmans, and Mooij}}]{lupacscu2006high}
\bibinfo{author}{\bibfnamefont{A.}~\bibnamefont{Lupa{\c{s}}cu}},
  \bibinfo{author}{\bibfnamefont{E.}~\bibnamefont{Driessen}},
  \bibinfo{author}{\bibfnamefont{L.}~\bibnamefont{Roschier}},
  \bibinfo{author}{\bibfnamefont{C.}~\bibnamefont{Harmans}}, \bibnamefont{and}
  \bibinfo{author}{\bibfnamefont{J.}~\bibnamefont{Mooij}},
  \bibinfo{journal}{Phys. Rev. Lett.} \textbf{\bibinfo{volume}{96}},
  \bibinfo{pages}{127003} (\bibinfo{year}{2006}).

\bibitem[{\citenamefont{Paauw et~al.}(2009)\citenamefont{Paauw, Fedorov,
  Harmans, and Mooij}}]{paauw2009tuning}
\bibinfo{author}{\bibfnamefont{F.}~\bibnamefont{Paauw}},
  \bibinfo{author}{\bibfnamefont{A.}~\bibnamefont{Fedorov}},
  \bibinfo{author}{\bibfnamefont{C.~M.} \bibnamefont{Harmans}},
  \bibnamefont{and} \bibinfo{author}{\bibfnamefont{J.}~\bibnamefont{Mooij}},
  \bibinfo{journal}{Phys. Rev. Lett.} \textbf{\bibinfo{volume}{102}},
  \bibinfo{pages}{090501} (\bibinfo{year}{2009}).

\bibitem[{\citenamefont{Lanting et~al.}(2014)\citenamefont{Lanting, Przybysz,
  Smirnov, Spedalieri, Amin, Berkley, Harris, Altomare, Boixo, Bunyk
  et~al.}}]{Lanting2014EntanglentInFluxQubit}
\bibinfo{author}{\bibfnamefont{T.}~\bibnamefont{Lanting}},
  \bibinfo{author}{\bibfnamefont{A.~J.} \bibnamefont{Przybysz}},
  \bibinfo{author}{\bibfnamefont{A.~Y.} \bibnamefont{Smirnov}},
  \bibinfo{author}{\bibfnamefont{F.~M.} \bibnamefont{Spedalieri}},
  \bibinfo{author}{\bibfnamefont{M.~H.} \bibnamefont{Amin}},
  \bibinfo{author}{\bibfnamefont{A.~J.} \bibnamefont{Berkley}},
  \bibinfo{author}{\bibfnamefont{R.}~\bibnamefont{Harris}},
  \bibinfo{author}{\bibfnamefont{F.}~\bibnamefont{Altomare}},
  \bibinfo{author}{\bibfnamefont{S.}~\bibnamefont{Boixo}},
  \bibinfo{author}{\bibfnamefont{P.}~\bibnamefont{Bunyk}},
  \bibnamefont{et~al.}, \bibinfo{journal}{Phys. Rev. X}
  \textbf{\bibinfo{volume}{4}}, \bibinfo{pages}{021041} (\bibinfo{year}{2014}).

\bibitem[{\citenamefont{You et~al.}(2007)\citenamefont{You, Hu, Ashhab, and
  Nori}}]{you2007c-shunttheory}
\bibinfo{author}{\bibfnamefont{J.~Q.} \bibnamefont{You}},
  \bibinfo{author}{\bibfnamefont{X.}~\bibnamefont{Hu}},
  \bibinfo{author}{\bibfnamefont{S.}~\bibnamefont{Ashhab}}, \bibnamefont{and}
  \bibinfo{author}{\bibfnamefont{F.}~\bibnamefont{Nori}},
  \bibinfo{journal}{Phys. Rev. B} \textbf{\bibinfo{volume}{75}}
  (\bibinfo{year}{2007}).

\bibitem[{\citenamefont{Lupa\c{s}cu et~al.}(2007)\citenamefont{Lupa\c{s}cu,
  Saito, Picot, de~Groot, Harmans, and Mooij}}]{lupascu2007DelftJBA}
\bibinfo{author}{\bibfnamefont{A.}~\bibnamefont{Lupa\c{s}cu}},
  \bibinfo{author}{\bibfnamefont{S.}~\bibnamefont{Saito}},
  \bibinfo{author}{\bibfnamefont{T.}~\bibnamefont{Picot}},
  \bibinfo{author}{\bibfnamefont{P.~C.} \bibnamefont{de~Groot}},
  \bibinfo{author}{\bibfnamefont{C.~J. P.~M.} \bibnamefont{Harmans}},
  \bibnamefont{and} \bibinfo{author}{\bibfnamefont{J.~E.} \bibnamefont{Mooij}},
  \bibinfo{journal}{Nature Physics} \textbf{\bibinfo{volume}{3}}
  (\bibinfo{year}{2007}).

\bibitem[{\citenamefont{Bal et~al.}(2012)\citenamefont{Bal, Deng, Orgiazzi,
  Ong, and Lupascu}}]{bal2012ultrasensitive}
\bibinfo{author}{\bibfnamefont{M.}~\bibnamefont{Bal}},
  \bibinfo{author}{\bibfnamefont{C.}~\bibnamefont{Deng}},
  \bibinfo{author}{\bibfnamefont{J.-L.} \bibnamefont{Orgiazzi}},
  \bibinfo{author}{\bibfnamefont{F.}~\bibnamefont{Ong}}, \bibnamefont{and}
  \bibinfo{author}{\bibfnamefont{A.}~\bibnamefont{Lupascu}},
  \bibinfo{journal}{Nat. Commun.} \textbf{\bibinfo{volume}{3}},
  \bibinfo{pages}{1324} (\bibinfo{year}{2012}).

\bibitem[{\citenamefont{Marcos et~al.}(2010)\citenamefont{Marcos, Wubs, Taylor,
  Aguado, Lukin, and S{\o}rensen}}]{Marcos2010couplingNVtoFQ}
\bibinfo{author}{\bibfnamefont{D.}~\bibnamefont{Marcos}},
  \bibinfo{author}{\bibfnamefont{M.}~\bibnamefont{Wubs}},
  \bibinfo{author}{\bibfnamefont{J.~M.} \bibnamefont{Taylor}},
  \bibinfo{author}{\bibfnamefont{R.}~\bibnamefont{Aguado}},
  \bibinfo{author}{\bibfnamefont{M.~D.} \bibnamefont{Lukin}}, \bibnamefont{and}
  \bibinfo{author}{\bibfnamefont{A.~S.} \bibnamefont{S{\o}rensen}},
  \bibinfo{journal}{Phys. Rev. Lett.} \textbf{\bibinfo{volume}{105}},
  \bibinfo{pages}{210501} (\bibinfo{year}{2010}).

\bibitem[{\citenamefont{Twamley and
  Barrett}(2010)}]{twamley2010superconducting}
\bibinfo{author}{\bibfnamefont{J.}~\bibnamefont{Twamley}} \bibnamefont{and}
  \bibinfo{author}{\bibfnamefont{S.~D.} \bibnamefont{Barrett}},
  \bibinfo{journal}{Phys. Rev. B} \textbf{\bibinfo{volume}{81}},
  \bibinfo{pages}{241202} (\bibinfo{year}{2010}).

\bibitem[{\citenamefont{Zhu et~al.}(2011)\citenamefont{Zhu, Saito, Kemp,
  Kakuyanagi, Karimoto, Nakano, Munro, Tokura, Everitt, Nemoto
  et~al.}}]{zhu2011coherentcoupling}
\bibinfo{author}{\bibfnamefont{X.}~\bibnamefont{Zhu}},
  \bibinfo{author}{\bibfnamefont{S.}~\bibnamefont{Saito}},
  \bibinfo{author}{\bibfnamefont{A.}~\bibnamefont{Kemp}},
  \bibinfo{author}{\bibfnamefont{K.}~\bibnamefont{Kakuyanagi}},
  \bibinfo{author}{\bibfnamefont{S.-i.} \bibnamefont{Karimoto}},
  \bibinfo{author}{\bibfnamefont{H.}~\bibnamefont{Nakano}},
  \bibinfo{author}{\bibfnamefont{W.~J.} \bibnamefont{Munro}},
  \bibinfo{author}{\bibfnamefont{Y.}~\bibnamefont{Tokura}},
  \bibinfo{author}{\bibfnamefont{M.~S.} \bibnamefont{Everitt}},
  \bibinfo{author}{\bibfnamefont{K.}~\bibnamefont{Nemoto}},
  \bibnamefont{et~al.}, \bibinfo{journal}{Nature}
  \textbf{\bibinfo{volume}{478}} (\bibinfo{year}{2011}).

\bibitem[{\citenamefont{Saito et~al.}(2013)\citenamefont{Saito, Zhu,
  Ams\"{u}ss, Matsuzaki, Kakuyanagi, Shimo-Oka, Mizuochi, Nemoto, Munro, and
  Semba}}]{shiro2013quantummemorySQ}
\bibinfo{author}{\bibfnamefont{S.}~\bibnamefont{Saito}},
  \bibinfo{author}{\bibfnamefont{X.}~\bibnamefont{Zhu}},
  \bibinfo{author}{\bibfnamefont{R.}~\bibnamefont{Ams\"{u}ss}},
  \bibinfo{author}{\bibfnamefont{Y.}~\bibnamefont{Matsuzaki}},
  \bibinfo{author}{\bibfnamefont{K.}~\bibnamefont{Kakuyanagi}},
  \bibinfo{author}{\bibfnamefont{T.}~\bibnamefont{Shimo-Oka}},
  \bibinfo{author}{\bibfnamefont{N.}~\bibnamefont{Mizuochi}},
  \bibinfo{author}{\bibfnamefont{K.}~\bibnamefont{Nemoto}},
  \bibinfo{author}{\bibfnamefont{W.~J.} \bibnamefont{Munro}}, \bibnamefont{and}
  \bibinfo{author}{\bibfnamefont{K.}~\bibnamefont{Semba}},
  \bibinfo{journal}{Phys. Rev. Lett.} \textbf{\bibinfo{volume}{111}},
  \bibinfo{pages}{107008} (\bibinfo{year}{2013}).

\bibitem[{\citenamefont{Matsuzaki
  et~al.}(2015{\natexlab{a}})\citenamefont{Matsuzaki, Zhu, Kakuyanagi, Toida,
  Shimooka, Mizuochi, Nemoto, Semba, Munro, Yamaguchi
  et~al.}}]{pramatsuzaki2015improving}
\bibinfo{author}{\bibfnamefont{Y.}~\bibnamefont{Matsuzaki}},
  \bibinfo{author}{\bibfnamefont{X.}~\bibnamefont{Zhu}},
  \bibinfo{author}{\bibfnamefont{K.}~\bibnamefont{Kakuyanagi}},
  \bibinfo{author}{\bibfnamefont{H.}~\bibnamefont{Toida}},
  \bibinfo{author}{\bibfnamefont{T.}~\bibnamefont{Shimooka}},
  \bibinfo{author}{\bibfnamefont{N.}~\bibnamefont{Mizuochi}},
  \bibinfo{author}{\bibfnamefont{K.}~\bibnamefont{Nemoto}},
  \bibinfo{author}{\bibfnamefont{K.}~\bibnamefont{Semba}},
  \bibinfo{author}{\bibfnamefont{W.}~\bibnamefont{Munro}},
  \bibinfo{author}{\bibfnamefont{H.}~\bibnamefont{Yamaguchi}},
  \bibnamefont{et~al.}, \bibinfo{journal}{Phys. Rev. A}
  \textbf{\bibinfo{volume}{91}}, \bibinfo{pages}{042329}
  (\bibinfo{year}{2015}{\natexlab{a}}).

\bibitem[{\citenamefont{Matsuzaki
  et~al.}(2015{\natexlab{b}})\citenamefont{Matsuzaki, Zhu, Kakuyanagi, Toida,
  Shimo-Oka, Mizuochi, Nemoto, Semba, Munro, Yamaguchi
  et~al.}}]{matsuzaki2015improving}
\bibinfo{author}{\bibfnamefont{Y.}~\bibnamefont{Matsuzaki}},
  \bibinfo{author}{\bibfnamefont{X.}~\bibnamefont{Zhu}},
  \bibinfo{author}{\bibfnamefont{K.}~\bibnamefont{Kakuyanagi}},
  \bibinfo{author}{\bibfnamefont{H.}~\bibnamefont{Toida}},
  \bibinfo{author}{\bibfnamefont{T.}~\bibnamefont{Shimo-Oka}},
  \bibinfo{author}{\bibfnamefont{N.}~\bibnamefont{Mizuochi}},
  \bibinfo{author}{\bibfnamefont{K.}~\bibnamefont{Nemoto}},
  \bibinfo{author}{\bibfnamefont{K.}~\bibnamefont{Semba}},
  \bibinfo{author}{\bibfnamefont{W.~J.} \bibnamefont{Munro}},
  \bibinfo{author}{\bibfnamefont{H.}~\bibnamefont{Yamaguchi}},
  \bibnamefont{et~al.}, \bibinfo{journal}{Phys. Rev. Lett.}
  \textbf{\bibinfo{volume}{114}}, \bibinfo{pages}{120501}
  (\bibinfo{year}{2015}{\natexlab{b}}).

\bibitem[{\citenamefont{Kubo et~al.}(2011)\citenamefont{Kubo, Grezes, Dewes,
  Umeda, Isoya, Sumiya, Morishita, Abe, Onoda, Ohshima
  et~al.}}]{kubo2011hybrid}
\bibinfo{author}{\bibfnamefont{Y.}~\bibnamefont{Kubo}},
  \bibinfo{author}{\bibfnamefont{C.}~\bibnamefont{Grezes}},
  \bibinfo{author}{\bibfnamefont{A.}~\bibnamefont{Dewes}},
  \bibinfo{author}{\bibfnamefont{T.}~\bibnamefont{Umeda}},
  \bibinfo{author}{\bibfnamefont{J.}~\bibnamefont{Isoya}},
  \bibinfo{author}{\bibfnamefont{H.}~\bibnamefont{Sumiya}},
  \bibinfo{author}{\bibfnamefont{N.}~\bibnamefont{Morishita}},
  \bibinfo{author}{\bibfnamefont{H.}~\bibnamefont{Abe}},
  \bibinfo{author}{\bibfnamefont{S.}~\bibnamefont{Onoda}},
  \bibinfo{author}{\bibfnamefont{T.}~\bibnamefont{Ohshima}},
  \bibnamefont{et~al.}, \bibinfo{journal}{Phys. Rev. Lett.}
  \textbf{\bibinfo{volume}{107}}, \bibinfo{pages}{220501}
  (\bibinfo{year}{2011}).

\bibitem[{\citenamefont{Matsuzaki and Nakano}(2012)}]{matsuzaki2012enhanced}
\bibinfo{author}{\bibfnamefont{Y.}~\bibnamefont{Matsuzaki}} \bibnamefont{and}
  \bibinfo{author}{\bibfnamefont{H.}~\bibnamefont{Nakano}},
  \bibinfo{journal}{Phys. Rev. B} \textbf{\bibinfo{volume}{86}},
  \bibinfo{pages}{184501} (\bibinfo{year}{2012}).

\bibitem[{\citenamefont{Tanaka et~al.}(2015)\citenamefont{Tanaka, Knott,
  Matsuzaki, Dooley, Yamaguchi, Munro, and Saito}}]{tanaka2015proposed}
\bibinfo{author}{\bibfnamefont{T.}~\bibnamefont{Tanaka}},
  \bibinfo{author}{\bibfnamefont{P.}~\bibnamefont{Knott}},
  \bibinfo{author}{\bibfnamefont{Y.}~\bibnamefont{Matsuzaki}},
  \bibinfo{author}{\bibfnamefont{S.}~\bibnamefont{Dooley}},
  \bibinfo{author}{\bibfnamefont{H.}~\bibnamefont{Yamaguchi}},
  \bibinfo{author}{\bibfnamefont{W.~J.} \bibnamefont{Munro}}, \bibnamefont{and}
  \bibinfo{author}{\bibfnamefont{S.}~\bibnamefont{Saito}},
  \bibinfo{journal}{Phys. Rev. Lett.} \textbf{\bibinfo{volume}{115}},
  \bibinfo{pages}{170801} (\bibinfo{year}{2015}).

\bibitem[{\citenamefont{Dooley et~al.}(2016)\citenamefont{Dooley, Yukawa,
  Matsuzaki, Knee, Munro, and Nemoto}}]{dooley2016hybrid}
\bibinfo{author}{\bibfnamefont{S.}~\bibnamefont{Dooley}},
  \bibinfo{author}{\bibfnamefont{E.}~\bibnamefont{Yukawa}},
  \bibinfo{author}{\bibfnamefont{Y.}~\bibnamefont{Matsuzaki}},
  \bibinfo{author}{\bibfnamefont{G.~C.} \bibnamefont{Knee}},
  \bibinfo{author}{\bibfnamefont{W.~J.} \bibnamefont{Munro}}, \bibnamefont{and}
  \bibinfo{author}{\bibfnamefont{K.}~\bibnamefont{Nemoto}},
  \bibinfo{journal}{New Journal of Physics} \textbf{\bibinfo{volume}{18}},
  \bibinfo{pages}{053011} (\bibinfo{year}{2016}).

\bibitem[{\citenamefont{Schuster et~al.}(2010)\citenamefont{Schuster, Sears,
  Ginossar, DiCarlo, Frunzio, Morton, Wu, Briggs, Buckley, Awschalom
  et~al.}}]{schuster2010high}
\bibinfo{author}{\bibfnamefont{D.~I.} \bibnamefont{Schuster}},
  \bibinfo{author}{\bibfnamefont{A.~P.} \bibnamefont{Sears}},
  \bibinfo{author}{\bibfnamefont{E.}~\bibnamefont{Ginossar}},
  \bibinfo{author}{\bibfnamefont{L.}~\bibnamefont{DiCarlo}},
  \bibinfo{author}{\bibfnamefont{L.}~\bibnamefont{Frunzio}},
  \bibinfo{author}{\bibfnamefont{J.~J.~L.} \bibnamefont{Morton}},
  \bibinfo{author}{\bibfnamefont{H.}~\bibnamefont{Wu}},
  \bibinfo{author}{\bibfnamefont{G.~A.~D.} \bibnamefont{Briggs}},
  \bibinfo{author}{\bibfnamefont{B.~B.} \bibnamefont{Buckley}},
  \bibinfo{author}{\bibfnamefont{D.~D.} \bibnamefont{Awschalom}},
  \bibnamefont{et~al.}, \bibinfo{journal}{Phys. Rev. Lett.}
  \textbf{\bibinfo{volume}{105}}, \bibinfo{pages}{140501}
  (\bibinfo{year}{2010}).

\bibitem[{\citenamefont{Kubo et~al.}(2010)\citenamefont{Kubo, Ong, Bertet,
  Vion, Jacques, Zheng, Dr{\'e}au, Roch, Auffeves, Jelezko
  et~al.}}]{kubo2010strong}
\bibinfo{author}{\bibfnamefont{Y.}~\bibnamefont{Kubo}},
  \bibinfo{author}{\bibfnamefont{F.~R.} \bibnamefont{Ong}},
  \bibinfo{author}{\bibfnamefont{P.}~\bibnamefont{Bertet}},
  \bibinfo{author}{\bibfnamefont{D.}~\bibnamefont{Vion}},
  \bibinfo{author}{\bibfnamefont{V.}~\bibnamefont{Jacques}},
  \bibinfo{author}{\bibfnamefont{D.}~\bibnamefont{Zheng}},
  \bibinfo{author}{\bibfnamefont{A.}~\bibnamefont{Dr{\'e}au}},
  \bibinfo{author}{\bibfnamefont{J.~F.} \bibnamefont{Roch}},
  \bibinfo{author}{\bibfnamefont{A.}~\bibnamefont{Auffeves}},
  \bibinfo{author}{\bibfnamefont{F.}~\bibnamefont{Jelezko}},
  \bibnamefont{et~al.}, \bibinfo{journal}{Phys. Rev. Lett.}
  \textbf{\bibinfo{volume}{105}}, \bibinfo{pages}{140502}
  (\bibinfo{year}{2010}).

\bibitem[{\citenamefont{Kubo et~al.}(2012)\citenamefont{Kubo, Diniz, Grezes,
  Umeda, Isoya, Sumiya, Yamamoto, Abe, Onoda, Ohshima
  et~al.}}]{kubo2012electron}
\bibinfo{author}{\bibfnamefont{Y.}~\bibnamefont{Kubo}},
  \bibinfo{author}{\bibfnamefont{I.}~\bibnamefont{Diniz}},
  \bibinfo{author}{\bibfnamefont{C.}~\bibnamefont{Grezes}},
  \bibinfo{author}{\bibfnamefont{T.}~\bibnamefont{Umeda}},
  \bibinfo{author}{\bibfnamefont{J.}~\bibnamefont{Isoya}},
  \bibinfo{author}{\bibfnamefont{H.}~\bibnamefont{Sumiya}},
  \bibinfo{author}{\bibfnamefont{T.}~\bibnamefont{Yamamoto}},
  \bibinfo{author}{\bibfnamefont{H.}~\bibnamefont{Abe}},
  \bibinfo{author}{\bibfnamefont{S.}~\bibnamefont{Onoda}},
  \bibinfo{author}{\bibfnamefont{T.}~\bibnamefont{Ohshima}},
  \bibnamefont{et~al.}, \bibinfo{journal}{Phys. Rev. B}
  \textbf{\bibinfo{volume}{86}}, \bibinfo{pages}{064514}
  (\bibinfo{year}{2012}).

\bibitem[{\citenamefont{Bienfait et~al.}(2016)\citenamefont{Bienfait, Pla,
  Kubo, Stern, Zhou, Lo, Weis, Schenkel, Thewalt, Vion
  et~al.}}]{bienfait2016reaching}
\bibinfo{author}{\bibfnamefont{A.}~\bibnamefont{Bienfait}},
  \bibinfo{author}{\bibfnamefont{J.}~\bibnamefont{Pla}},
  \bibinfo{author}{\bibfnamefont{Y.}~\bibnamefont{Kubo}},
  \bibinfo{author}{\bibfnamefont{M.}~\bibnamefont{Stern}},
  \bibinfo{author}{\bibfnamefont{X.}~\bibnamefont{Zhou}},
  \bibinfo{author}{\bibfnamefont{C.}~\bibnamefont{Lo}},
  \bibinfo{author}{\bibfnamefont{C.}~\bibnamefont{Weis}},
  \bibinfo{author}{\bibfnamefont{T.}~\bibnamefont{Schenkel}},
  \bibinfo{author}{\bibfnamefont{M.}~\bibnamefont{Thewalt}},
  \bibinfo{author}{\bibfnamefont{D.}~\bibnamefont{Vion}}, \bibnamefont{et~al.},
  \bibinfo{journal}{Nature nanotechnology} \textbf{\bibinfo{volume}{11}},
  \bibinfo{pages}{253} (\bibinfo{year}{2016}).

\bibitem[{\citenamefont{Probst et~al.}(2017)\citenamefont{Probst, Bienfait,
  Campagne-Ibarcq, Pla, Albanese, Da~Silva~Barbosa, Schenkel, Vion, Esteve,
  M{\o}lmer et~al.}}]{probst2017inductive}
\bibinfo{author}{\bibfnamefont{S.}~\bibnamefont{Probst}},
  \bibinfo{author}{\bibfnamefont{A.}~\bibnamefont{Bienfait}},
  \bibinfo{author}{\bibfnamefont{P.}~\bibnamefont{Campagne-Ibarcq}},
  \bibinfo{author}{\bibfnamefont{J.}~\bibnamefont{Pla}},
  \bibinfo{author}{\bibfnamefont{B.}~\bibnamefont{Albanese}},
  \bibinfo{author}{\bibfnamefont{J.}~\bibnamefont{Da~Silva~Barbosa}},
  \bibinfo{author}{\bibfnamefont{T.}~\bibnamefont{Schenkel}},
  \bibinfo{author}{\bibfnamefont{D.}~\bibnamefont{Vion}},
  \bibinfo{author}{\bibfnamefont{D.}~\bibnamefont{Esteve}},
  \bibinfo{author}{\bibfnamefont{K.}~\bibnamefont{M{\o}lmer}},
  \bibnamefont{et~al.}, \bibinfo{journal}{Applied Physics Letters}
  \textbf{\bibinfo{volume}{111}}, \bibinfo{pages}{202604}
  (\bibinfo{year}{2017}).

\bibitem[{\citenamefont{Toida et~al.}(2019)\citenamefont{Toida, Matsuzaki,
  Kakuyanagi, Zhu, Munro, Yamaguchi, and Saito}}]{toida2017ESRwithFQ}
\bibinfo{author}{\bibfnamefont{H.}~\bibnamefont{Toida}},
  \bibinfo{author}{\bibfnamefont{Y.}~\bibnamefont{Matsuzaki}},
  \bibinfo{author}{\bibfnamefont{K.}~\bibnamefont{Kakuyanagi}},
  \bibinfo{author}{\bibfnamefont{X.}~\bibnamefont{Zhu}},
  \bibinfo{author}{\bibfnamefont{W.~J.} \bibnamefont{Munro}},
  \bibinfo{author}{\bibfnamefont{H.}~\bibnamefont{Yamaguchi}},
  \bibnamefont{and} \bibinfo{author}{\bibfnamefont{S.}~\bibnamefont{Saito}},
  \bibinfo{journal}{Communications Physics} \textbf{\bibinfo{volume}{2}},
  \bibinfo{pages}{33} (\bibinfo{year}{2019}).

\bibitem[{\citenamefont{Angerer et~al.}(2017)\citenamefont{Angerer, Putz,
  Krimer, Astner, Zens, Glattauer, Streltsov, Munro, Nemoto, Rotter
  et~al.}}]{Angerere2017ultralongrelaxation}
\bibinfo{author}{\bibfnamefont{A.}~\bibnamefont{Angerer}},
  \bibinfo{author}{\bibfnamefont{S.}~\bibnamefont{Putz}},
  \bibinfo{author}{\bibfnamefont{D.~O.} \bibnamefont{Krimer}},
  \bibinfo{author}{\bibfnamefont{T.}~\bibnamefont{Astner}},
  \bibinfo{author}{\bibfnamefont{M.}~\bibnamefont{Zens}},
  \bibinfo{author}{\bibfnamefont{R.}~\bibnamefont{Glattauer}},
  \bibinfo{author}{\bibfnamefont{K.}~\bibnamefont{Streltsov}},
  \bibinfo{author}{\bibfnamefont{W.~J.} \bibnamefont{Munro}},
  \bibinfo{author}{\bibfnamefont{K.}~\bibnamefont{Nemoto}},
  \bibinfo{author}{\bibfnamefont{S.}~\bibnamefont{Rotter}},
  \bibnamefont{et~al.}, \bibinfo{journal}{Science Advances}
  \textbf{\bibinfo{volume}{3}} (\bibinfo{year}{2017}).

\bibitem[{\citenamefont{Budoyo et~al.}(2018)\citenamefont{Budoyo, Kakuyanagi,
  Toida, Matsuzaki, Munro, Yamaguchi, and Saito}}]{rangga2018phonon-bottleneck}
\bibinfo{author}{\bibfnamefont{R.~P.} \bibnamefont{Budoyo}},
  \bibinfo{author}{\bibfnamefont{K.}~\bibnamefont{Kakuyanagi}},
  \bibinfo{author}{\bibfnamefont{H.}~\bibnamefont{Toida}},
  \bibinfo{author}{\bibfnamefont{Y.}~\bibnamefont{Matsuzaki}},
  \bibinfo{author}{\bibfnamefont{W.~J.} \bibnamefont{Munro}},
  \bibinfo{author}{\bibfnamefont{H.}~\bibnamefont{Yamaguchi}},
  \bibnamefont{and} \bibinfo{author}{\bibfnamefont{S.}~\bibnamefont{Saito}},
  \bibinfo{journal}{Applied Physics Express} \textbf{\bibinfo{volume}{11}},
  \bibinfo{pages}{043002} (\bibinfo{year}{2018}).

\bibitem[{\citenamefont{Ams\"uss et~al.}(2011)\citenamefont{Ams\"uss, Koller,
  N\"obauer, Putz, Rotter, Sandner, Schneider, Schramb\"ock, Steinhauser,
  Ritsch et~al.}}]{amususs2011cavitycoupledspin}
\bibinfo{author}{\bibfnamefont{R.}~\bibnamefont{Ams\"uss}},
  \bibinfo{author}{\bibfnamefont{C.}~\bibnamefont{Koller}},
  \bibinfo{author}{\bibfnamefont{T.}~\bibnamefont{N\"obauer}},
  \bibinfo{author}{\bibfnamefont{S.}~\bibnamefont{Putz}},
  \bibinfo{author}{\bibfnamefont{S.}~\bibnamefont{Rotter}},
  \bibinfo{author}{\bibfnamefont{K.}~\bibnamefont{Sandner}},
  \bibinfo{author}{\bibfnamefont{S.}~\bibnamefont{Schneider}},
  \bibinfo{author}{\bibfnamefont{M.}~\bibnamefont{Schramb\"ock}},
  \bibinfo{author}{\bibfnamefont{G.}~\bibnamefont{Steinhauser}},
  \bibinfo{author}{\bibfnamefont{H.}~\bibnamefont{Ritsch}},
  \bibnamefont{et~al.}, \bibinfo{journal}{Phys. Rev. Lett.}
  \textbf{\bibinfo{volume}{107}}, \bibinfo{pages}{060502}
  (\bibinfo{year}{2011}).

\bibitem[{\citenamefont{Ajoy et~al.}(2017)\citenamefont{Ajoy, Liu, and
  Cappellaro}}]{ajoy2016DCmagnetmetry}
\bibinfo{author}{\bibfnamefont{A.}~\bibnamefont{Ajoy}},
  \bibinfo{author}{\bibfnamefont{Y.~X.} \bibnamefont{Liu}}, \bibnamefont{and}
  \bibinfo{author}{\bibfnamefont{P.}~\bibnamefont{Cappellaro}},
  \bibinfo{journal}{arXiv preprint arXiv:1611.04691}  (\bibinfo{year}{2017}).

\bibitem[{\citenamefont{Bylander et~al.}(2011)\citenamefont{Bylander,
  Gustavsson, Yan, Yoshihara, Harrabi, Fitch, Cory, Nakamura, Tsai, and
  Oliver}}]{jonas2011noisespectroscopywithFQ}
\bibinfo{author}{\bibfnamefont{J.}~\bibnamefont{Bylander}},
  \bibinfo{author}{\bibfnamefont{S.}~\bibnamefont{Gustavsson}},
  \bibinfo{author}{\bibfnamefont{F.}~\bibnamefont{Yan}},
  \bibinfo{author}{\bibfnamefont{F.}~\bibnamefont{Yoshihara}},
  \bibinfo{author}{\bibfnamefont{K.}~\bibnamefont{Harrabi}},
  \bibinfo{author}{\bibfnamefont{G.}~\bibnamefont{Fitch}},
  \bibinfo{author}{\bibfnamefont{D.~G.} \bibnamefont{Cory}},
  \bibinfo{author}{\bibfnamefont{Y.}~\bibnamefont{Nakamura}},
  \bibinfo{author}{\bibfnamefont{J.-S.} \bibnamefont{Tsai}}, \bibnamefont{and}
  \bibinfo{author}{\bibfnamefont{W.~D.} \bibnamefont{Oliver}},
  \bibinfo{journal}{Nature Physics} \textbf{\bibinfo{volume}{7}},
  \bibinfo{pages}{565} (\bibinfo{year}{2011}).

\bibitem[{\citenamefont{Kitazawa et~al.}(2017)\citenamefont{Kitazawa,
  Matsuzaki, Saijo, Kakuyanagi, Saito, and
  Ishi-Hayase}}]{kitazawa2017vectormagnetsensing}
\bibinfo{author}{\bibfnamefont{S.}~\bibnamefont{Kitazawa}},
  \bibinfo{author}{\bibfnamefont{Y.}~\bibnamefont{Matsuzaki}},
  \bibinfo{author}{\bibfnamefont{S.}~\bibnamefont{Saijo}},
  \bibinfo{author}{\bibfnamefont{K.}~\bibnamefont{Kakuyanagi}},
  \bibinfo{author}{\bibfnamefont{S.}~\bibnamefont{Saito}}, \bibnamefont{and}
  \bibinfo{author}{\bibfnamefont{J.}~\bibnamefont{Ishi-Hayase}},
  \bibinfo{journal}{Phys. Rev. A} \textbf{\bibinfo{volume}{96}},
  \bibinfo{pages}{042115} (\bibinfo{year}{2017}).

\bibitem[{\citenamefont{Subramanian et~al.}(1998)\citenamefont{Subramanian,
  Lam, and Webb}}]{subramanian1998RFmicrocoilNMR}
\bibinfo{author}{\bibfnamefont{R.}~\bibnamefont{Subramanian}},
  \bibinfo{author}{\bibfnamefont{M.~M.} \bibnamefont{Lam}}, \bibnamefont{and}
  \bibinfo{author}{\bibfnamefont{A.~G.} \bibnamefont{Webb}},
  \bibinfo{journal}{Journal of Magnetic Resonance}
  \textbf{\bibinfo{volume}{133}}, \bibinfo{pages}{227} (\bibinfo{year}{1998}).

\end{thebibliography}
\end{document}